\documentclass[AMA,STIX1COL]{WileyNJD-v2}

\articletype{research article}%
\usepackage{threeparttable}
\received{26 April 2016}
\revised{6 June 2016}
\accepted{6 June 2016}

\begin{document}
	
	\title{Robust statistical inference for the matched net benefit and  the matched win ratio using prioritized composite endpoints}
	\author[1,2]{Roland A. Matsouaka*}
	\author[3]{Adrian Coles}
	
	\authormark{MATSOUAKA \textsc{et al}}
	
\address[1]{\orgdiv{Department of Biostatistics and Bioinformatics}, \orgname{Duke University}, \orgaddress{\state{Durham}, \country{North Carolina, USA}}}
\address[2]{\orgdiv{Program for Comparative Effectiveness Methodology}, \orgname{Duke Clinical Research Institute, Duke University}, \orgaddress{\state{Durham}, \country{North Carolina, USA}}}
\address[3]{\orgdiv{Lilly Research Laboratories}, \orgname{Eli Lilly and Company}, \orgaddress{\state{Indianapolis}, \country{Indiana, USA}} (adrian.coles@lilly.com)}

\corres{*Roland A. Matsouaka\\ Duke Clinical Research Institute, \\6021 North Pavilion, 2400 Pratt Street,  Durham, NC 27705, USA\\\email{roland.matsouaka@duke.edu}}	
%
%
%
	
	
		\abstract[Abstract]{

		As alternatives to the  time-to-first-event analysis of composite endpoints, the  {\it net benefit} (NB) and   the {\it win ratio} (WR)---which assess treatment effects using prioritized component  outcomes based on clinical importance--- have been proposed.  However,  statistical inference of NB and WR relies on a large-sample assumptions, which can lead to an invalid test statistic and inadequate, unsatisfactory confidence intervals, especially when the sample size is small or the proportion of wins is near  0 or 1.
		
		In this paper, we  develop  a systematic approach to address these limitations in a paired-sample design. We first introduce a new test statistic under the null hypothesis of no treatment difference.  Then, we present the formula to calculate the  sample size. Finally, we develop the confidence interval estimations of these two estimators.  To estimate the confidence intervals, we use the {\it method of variance estimates recovery} (MOVER), that combines two separate individual-proportion confidence intervals into a hybrid interval for the estimand of interest. We assess the performance of the proposed test statistic and MOVER confidence interval estimations through simulation studies. 		
		We demonstrate that the MOVER confidence intervals are  as good as the large-sample confidence intervals when the sample is large and when the proportions of wins is bounded away from 0 and 1. Moreover, the MOVER intervals outperform  their competitors when the sample is small or the proportions are at or near the boundaries 0 and 1.  We illustrate the method (and its competitors) using three examples from randomized clinical studies.}
	
	
	
	\maketitle
	

\section{Introduction}
Composite endpoints have become ubiquitous in biomedical research. They incorporate multiple individual outcomes that make use of important clinical events and account for different morbidities that impact affect a patient's disease experience. A well-crafted composite endpoint which has the advantage to increase statistical precision and efficiency and often require smaller sample sizes; eliminate the need for multiple adjustment of outcome comparisons;  use fewer financial, human, and logistic resources; and expedite access to the study results. The commonly-used approach of time-to-first event handles all components of the composite endpoints as if they were of equal importance, focuses on a patient's first experienced event, and ignores subsequent events regardless of their impact. However, outcomes of lesser importance are more frequent and sometimes dominate the study finding, which may lead to a difficult understanding of the study results and their interpretation.\cite{matsouaka2018optimal}

As an alternative to the composite endpoint of time-to-first event, Pocock and colleagues introduced the win ratio  where  the component outcomes of interest are ordered based on a pre-specified hierarchy from the most severe to the least severe outcome and then pairwise comparisons of patients are considered.\cite{pocock2011win} With the matched win ratio, each patient from the treatment group is first matched to another patient in the control group based on their baseline risk profiles and time of enrollment.  Then, their respective outcomes are compared following the pre-specified hierarchy. For each pair, it is determined whether the patient in the treatment group is a winner or a loser depending on whether or not the treated patient has the more favorable outcome, starting first with the most severe outcome. If it cannot be determined whether the treated patient is a winner or a loser (there is a draw or patients cannot be compared because of censoring), then the pair is compared subsequently on the second most severe outcome and sequentially on the other outcomes following the pre-specify hierarchy. 
Finally, if no one is declared a winner by the last outcome in the hierarchy, the pair is counted as a tie. 

The matched win ratio comparing the treatment and the control groups is then the ratio of the number of matched pairs where treated patients had the more favorable outcome to the number of matched pairs where the treated patient fared worse.\cite{pocock2011win} Similarly, the net benefit (also known as the proportion in favor of treatment \cite{buyse2010generalized}) is the difference between the proportion of winners and the proportion of losers.\cite{verbeeck2020evaluation, luo2017weighted}
In this paper, we show how to test for the null hypothesis of no treatment difference and propose confidence interval formulas for both the matched win ratio and net benefit. Furthermore, as it is the case for the tandem risk ratio--risk difference in  their relationship to the disease assessment \cite{feng2019relations}, the win ratio and the net benefit are also different and provide different perspectives for assessing the magnitude of treatment effect on the disease under study. 

There is a close connection between hypothesis testing and confidence interval estimation. As Rosner argues, these are two distinct procedures aiming at providing a well-rounded inference of an estimand of interest.\cite{rosner2015fundamentals} Hypothesis testing via its p-value demonstrates precisely how significant is an estimate from the null hypothesis, whereas  confidence intervals provide additional information by giving a range of values within which the estimand is likely to fall into. This can have important  implications and help put the results from a study in a broader subject matter context. Thus the need to provide both the significance levels and confidence intervals as complementary information.\cite{rosner2015fundamentals} 

For the win ratio, Pocock et al. give a test statistic and a closed-form, large-sample confidence interval for the matched version of the win ratio.\cite{pocock2011win}  However, the test statistic is constructed using an estimate of standard error under the alternative hypothesis rather than the null hypothesis. While such a test statistic based on an estimate is (asymptotically) normally distributed, in small to moderate sample, its magnitude may be inflated and thus results in a type I error that is higher than the desired significance level and lead to unnecessarily over-optimistic results. Moreover, the confidence interval for the matched win ratio proposed by Pocock et al.\cite{pocock2011win} is a Wald-type binomial confidence interval, based on the normal approximation of a binomial distribution, and therefore inherits well-known limitations of the Wald intervals, including the potential for poor coverage in studies with small or moderate sample sizes or when the proportion of wins is near or at either of the boundaries 0 or 1.  Instead, a large-sample confidence interval based on a log transformation is preferred \cite{lachin2009biostatistical} or even better one may leverage novel optimal estimation methods that use hybrid approaches.\cite{newcombe2012confidence,newcombe1998improved,zou2008construction,donner2012closed} On the other, although it is straightforward to estimate the matched net benefit, this has never been done in the literature. 
Therefore, a systematically sound method  to estimate the matched net benefit and its corresponding confidence interval is needed.

Both hypothesis testing and confidence interval estimation for the matched net benefit and matched win ratio can be  seen from the lenses of statistical inference for the difference and ratio of proportions based on a paired data. In that perspective, we can develop a test that preserves the type I error and allows for reliable sample size estimation.. Moreover, several methods have been proposed to suitably derive  confidence intervals for differences and ratios of proportions that preserve boundary-respecting properties, sensible coverage probabilities, and better coverage-width trade-offs.

The goal of this paper is three-fold: propose a better test statistic for the hypothesis of no treatment difference,  provide a power and sample size formula, and construct confidence intervals for the paired win ratio and the net benefit. First, in Section \ref{sec:tssize}, we review the method presented by Pocock et al. \cite{pocock2011win} to highlight its limitation. Then, we propose  a new test statistic for the hypothesis of no treatment difference  as well as a formula to estimate the sample size for both the win ratio and the net benefit in the context of a paired-sample design. We also use the large-sample normal approximation of the multinomial distribution to estimate the large-sample confidence interval for the both the win ratio and the net benefit. Finally, we propose optimal alternative methods to estimate confidence for small and moderate sample sizes by inverting the test statistic under the null of no difference between treatments. 

The rest of this paper is organized as follows. 
In Section \ref{sec:conf_int},  we present methods for confidence interval estimations, starting with the net benefit in Section \ref{sec:win_diff} and then later the win ratio in Section \ref{sec:win_ratio}. For  both the win ratio and the net benefit, we begin with the large-sample normal approximation of the multinomial distribution to estimate their corresponding large-sample confidence intervals and  discuss their drawbacks.  We also introduce an alternative confidence interval for the win ratio based on the Fieller's theorem, in Section \ref{sec:fieller}. From the Fieller's theorem method, we give a brief summary of the pros and cons of using this method. 
We propose new alternatives to current confidence interval estimation, using the {\it method of variance estimates recovery} (MOVER) interval estimation method. We present a review of the MOVER in Section \ref{sec:wd_mover} and derive confidence intervals for the net benefit as well as the confidence intervals for the win ratio (in Section \ref{sec:mover_wr}).  We run and present the results of an extensive simulation study in Section \ref{sec:simulations}, with the main focus on the performance of the confidence intervals methods under small to moderate sample sizes. We use three examples, in Section \ref{sec:application}, to illustrate the different methods. Finally, in Section \ref{sec:conclusion}, we end the paper  with some concluding remarks and a recommendation.

\section{Tests statistics and sample size calculation}\label{sec:tssize}

\subsection{Notation}
We consider a study where  a new treatment is being evaluated against a   placebo treatment on a set of pre-specified, hierarchy of prioritized outcomes by comparing pairs of matched patients. Each pair is formed with one patient from the treatment group and one from the control group, based on specific criteria. Then, the paired patients are compared on their respective outcomes following the pre-specified hierarchy of outcomes. 

Starting from the most severe outcome, the pair is compared to determine whether the patient on the treatment group has a better outcome ("winner"), fared worse ("loser"), or there was no difference ("tie"). Whenever the treated patient wins or loses, we stop. Otherwise, the pair is compared on the next outcome of the hierarchy and so forth. If the entire hierarchy of outcomes is exhausted and there is no winner (or loser), the paired is then considered a tie. 

Consider $\pi_w, \pi_l, \pi_t$   the probabilities of wins, losses, and ties for the patients in the treatment group, our goal is to estimate the net benefit  $\text{NB}=\pi_w-\pi_l$  and the win ratio $\text{WR}=\displaystyle\frac{\pi_w}{\pi_l}$,  their corresponding confidence intervals, and the test statistic of no treatment effect $H_0: \pi_w=\pi_l$. 
Suppose that  we observe a total number of $N_w$ wins, $N_l$ losses, and $N_t$ ties in the treatment group, with  $N=N_w+N_l+N_t$ the total number of observed matched pairs. The proportions 
$\displaystyle p_w=\frac{N_w}{N}$, $\displaystyle p_l=\frac{N_l}{N}$, and $\displaystyle p_t=\frac{N_t}{N}$ are unbiased estimators of  $\pi_w, \pi_l$, and $\pi_t$  respectively. 
The net benefit and the win ratio are estimated, respectively, by $$\displaystyle D_w=\displaystyle\frac{N_w}{N}-\displaystyle\frac{N_l}{N}~~~~~\text{and}~~~~~\displaystyle R_w=N_w/N_l, ~\text{for} ~~N_l\ne 0.$$

As long as pairing patients is (considered as) an independent process where any variation in the matching process is considered negligible, the total number of matched pairs is (assumed to be) constant and fixed. In that respect the $N$ pairs of patients are considered conditionally independent from each other, conditional on the matching process that creates the pair. Hence, the random vector $(N_w, N_l, N_t)$ follow a multinomial distribution $M(N, \pi_w, \pi_l,\pi_t)$, with a  probability density function
\begin{eqnarray}\label{eq:multi}
f(N_w, N_l, N_t)=\frac{N!}{N_w! N_l! N_t!}\pi_w^{N_w}\pi_l^{N_l}\pi_t^{N_t}
\end{eqnarray} 
where $\pi_i$ is the probability of outcome $i$, i.e., win, loss, or tie (in short $w, l$, or $t$), such that  $\pi_w+\pi_l+ \pi_t=1$ and $N_w+ N_l+ N_t=N$. Therefore, $E(N_i)=Np_i$, $\text{Var}\, (N_i)=N\pi_i(1-\pi_i)$  and  $\text{Cov}(N_i, N_{j})= -N\pi_i\pi_{j},$ for $i,j=w, l, t$ and $i\ne j.$ \\By the multivariate central limit theorem, $\sqrt{N}\left[ ( p_w,  p_l,  p_t)-(\pi_w,  \pi_l,  \pi_t)\right]$ converges to a normal distribution with mean $(0, 0, 0)$ and covariance  matrix  $\Sigma=\left(\Sigma_{ij}\right)_{\substack{1\leq i, j\leq 3}}$ with  $\Sigma_{ii}=\pi_i(1-\pi_i)/N$  and  $\Sigma_{ij}= -\pi_i\pi_{j}/N,$ for $i,j=w, l, t$ and $i\ne j$.

Moreover, following Casela and Berger, \cite{casella2002statistical} we can show that the conditional probability mass function of the random  variable $N_w$ given $N_w+N_l=m$ is 
\begin{align}\label{eq:binomial}
f (N_w|N_w+N_l)=\displaystyle\frac{(N_w+N_l)!}{N_w!N_l!}\left( \frac{\pi_w}{\pi_w+\pi_l}\right) ^{N_w}\left( 1-\frac{\pi_w}{\pi_w+\pi_l}\right)^{N_l}
\end{align} 
This means, for fixed $N_w+N_l$, the random variable $N_w$ follows a binomial distribution with parameters $N_w+N_l$ and $Q=\displaystyle \frac{\pi_w}{\pi_w+\pi_l}.$
Such a distribution is particularly important to develop inference for the win ratio $R_w$, since one can define 
\begin{eqnarray}\label{eq:rpocock}
\displaystyle R_w=\frac{N_w/(N_w+N_l)}{N_l/(N_w+N_l)}
\end{eqnarray} 
and use the properties of the  probability mass function \eqref{eq:binomial} to derive the confidence interval for $R_w$ and use the conditional probability mass function of $N_w$ given $N_w+N_l.$ 
\subsection{Pocock et al.'s approach} \label{sec:pocock_short}
Pocock et al.\cite{pocock2011win} focused on the estimatiom of the win ratio and used the binomial probability mass function \eqref{eq:binomial} to determine its corresponding test statistic and confidence interval as we  show here. Their approach uses the normal approximation to the conditional binomial distribution \eqref{eq:binomial}, as we outline in this section. 

To estimate such a confidence interval for $R$, the first step is to  obtain the large-sample $100(1-\alpha)\%$ confidence interval $(Q_L, Q_U)$ of $Q=\displaystyle \frac{\pi_w}{\pi_w+\pi_l}$ by means of the central limit theorem as 
\begin{align}\label{eq:nus_binomial}
	Q_w\pm z_{\frac{\alpha}{2}}\sqrt{Q_w(1-Q_w)/(N_w+N_l)},
\end{align}  where $Q_w=\displaystyle \frac{p_w}{p_w+p_l}=\displaystyle \frac{N_w}{N_w+N_l}.$ Then, using the relationship $\displaystyle R_w=\frac{Q_w}{(1-Q_w)},$ we calculate the $100(1-\alpha)\%$ confidence interval  $
(R_L, R_U)$ of the win ratio $R$  as 
\begin{align}\label{eq:wr_binomial}
\displaystyle \left[ \frac{Q_L}{(1-Q_L)}~,~\frac{Q_U}{(1-Q_U)} \right].
\end{align}  
This is approach used by Pocock et al.\cite{pocock2011win} 

Moreover, the binomial distribution \eqref{eq:binomial} can be leverage to assess the null hypothesis of no difference between the two treatment groups $H_0: \pi_w=\pi_l$ against the two-sided alternative $H_1: \pi_w\ne \pi_l.$ Using the conditional probability $Q_w,$ the null and alternative hypothesis can also be expressed as 
$$H_0: Q_w=\frac{1}{2}~~\text{versus}~~ H_1: Q_w\neq \frac{1}{2}.$$ In that regard, Pocock et al. \cite{pocock2011win} suggest the standard normal deviate  
\begin{eqnarray}\label{eq:test_binomial}
Z_{P}=\frac{Q_w-0.5}{\sqrt{ Q_w(1-Q_w)/(N_w+N_l)}}
\end{eqnarray} 
as  test statistic for the null hypothesis $H_0: Q_w=\frac{1}{2}$, assuming that $Z_P$ follow a standard normal distribution. 

\subsection{Limitations and alternatives} 
As argued by Lachin \cite{lachin2009biostatistical} (pages 37-38), genuine tests statistic that are based on the standardized normal deviate are constructed using the estimated standard error $\widehat \sigma_0^2(Q_w)=1/4(N_w+N_l)$, derived under the null hypothesis $H_0$, not the estimate of the variance under the alternative hypothesis such as $\widehat\text{Var}\, (Q_w)=Q_w(1-Q_w)/(N_w+N_l).$ Otherwise, problems may arise as discussed by Mantel.\cite{mantel1987understanding} 
The fundamental reason is that  $\widehat \text{Var}\, (Q_w)\leq \widehat \sigma_0^2(Q_w)$ for any value of $Q_w$, i.e., the estimated variance under the null hypothesis $\widehat \sigma_0^2(Q_w)$ is underestimated when we replaced it by the estimated variance under the alternative  $\widehat \text{Var}\, (Q_w)$. Therefore,  a test statistic based on the estimated standard error  $\widehat \text{Var}\, (Q_w)$ is anti-conservative: it can inflate the magnitude of the test statistic $Z_{P}$ and drastically  increase the type I error, especially for small sample sizes.

Choosing a test statistic that is based on a variance  estimated under the null hypothesis, instead of the alternative hypothesis, gives a type I error probability that is closer to the desired significance level $\alpha$. Hence, as a correction, we recommend  using  the test statistic 
\begin{eqnarray}\label{eq:test_binom}
Z=\left(Q_w-0.5\right)\sqrt{4(N_w+N_l)}.
\end{eqnarray}
We should bear in mind that there is a perfect correspondence between the test of no net benefit, i.e., $H_0: p_w=p_l$ and the test of no win $H_0: \text{WR} =1$. Therefore, the above test statistic $Z$ can also be used for testing the  null hypothesis $H_0: \text{WR} =1$.
 
We can in fact make a direct connection between the wins, losses, ties of the win ratio method and the familiar discordance and concordance pairs used in paired-sample design. With such a similitude, the above test statistic $Z$ is simply the all too familiar McNemar's test statistic. 
This analogy with the paired-sample design also allows us to not recommend using  confidence interval \eqref{eq:wr_binomial} for the win ratio, especially for small or moderate sample sizes, for two specific reasons. First, $Q_w$ is the ratio of two probabilities and log transformations are preferred when estimating large-sample  confidence intervals for such ratios that direct estimation methods.\cite{gart1988approximate,lachin2009biostatistical} The reason is that log ratios tend to be closer to normal distributions than the ratio themselves, which leads to symmetrical or approximately symmetrical confidence intervals on the log scale. In fact, it is not uncommon to find adequately estimated bounds $Q_L$ and $Q_U$ of the ratio $Q_w$ that lead to an unsatisfactory confidence interval of the matched win ratio as we demonstrate in one of our data examples in Section \ref{sec:application_udca}.  Second, and most importantly, standard (Wald) large-sample confidence interval estimations of probabilities and of parameters comparing probabilities, although computationally simple and commonly used,  lead to confidence intervals with low coverage probability and that sometimes display erratic behaviors.\cite{agresti2005simple,agresti2000simple,newcombe2012confidence,brown2005confidence} 
A large-sample test statistic can be derived from the normal approximation to either the multinomial distribution \eqref{eq:multi} or the conditional binomial distribution \eqref{eq:binomial} as we show in Section \ref{sec:ls_test} below. 

While we can also use either the  probability mass function \eqref{eq:multi} or \eqref{eq:binomial} as basis to determine the test statistic as well as the point estimate and confidence intervals for the win ratio, we cannot  use the conditional probability mass function \eqref{eq:binomial} to derive  the necessary inference for the net benefit. In fact, we cannot use $Q_w$ and $Q_l=N_l/(N_w+N_l)$ to define the net benefit as $Q_w-Q_l$ since such an approach is only valid if $N_t+N_l$ is given (or fixed) in advance, i.e., if the number of ties is known in advance or we are preemptively sure that there are no ties at all. However, both scenarios are not reasonable and rarely happen in practice.  In the context of the win-loss comparisons inherent to estimation of the net benefit and the win ratio,  there is often a censoring for time-to-event outcomes, equality of or missing outcome values, etc. that lead to a probability of ties that is greater than 0. We provide in the Appendix some simulation results to demonstrate that using $Q_w-Q_l$, instead of  $(N_w-N_l)/N$, as the net benefit lead to bias when ties are present. Nevertheless, for either the win ratio or the net benefit, the test statistic \eqref{eq:test_binom} is a valid test statistic to the null hypothesis of no treatment difference.

Therefore, for the rest of this section, we will use the binomial distribution \eqref{eq:binomial} to derive the exact test statistic for the null hypothesis $H_0$. Then, we will derive the large sample test statistic using the multinomial distribution \eqref{eq:multi} and demonstrate that it is exactly equal to the above test statistic $Z$. Finally, we will use the test statistic to determine the formula  to estimate the  sample size in a study design where either the net benefit or the win ratio is the effect measure of interest. 
\subsection{Exact test statistic}\label{sec:e-test}
From the binomial distribution \eqref{eq:binomial}, under the null hypothesis $H_0,$ 
\begin{align*}
E(N_w|N_w+N_l)=\frac{N_w+N_l}{2}~~~\text{and}~~~ \text{Var}\,(N_w|N_w+N_l)=\frac{N_w+N_l}{4}
\end{align*}
In elementary Biostatistics courses (see for example Rosner \cite{rosner2015fundamentals}), the normal approximation to the binomial distribution \eqref{eq:binomial} will be valid only if $Var(N_w|N_w+N_l) \geq 5$, i.e, $(N_w+N_l)\geq 20.$ However, Samuels and Lu give a more comprehensive sets of guidelines for deciding when large is large enough \cite{samuels1992sample}. 

When such a condition is not satisfy,  a test based on exact binomial probability can be used. The exact 2-sided p-value is given by 
\begin{align*}
p= 
\begin{cases}
2\times\displaystyle\sum_{k=0}^{N_w}\begin{pmatrix}\displaystyle N_w+N_l\\  k\end{pmatrix}\left(\frac{1}{2} \right)^{N_w+N_l}       & ~ \text{if }~ N_w< N_l\\
2\times\displaystyle\sum_{k=N_w}^{N_w+N_l}\begin{pmatrix}\displaystyle N_w+N_l\\  k\end{pmatrix}\left(\frac{1}{2} \right)^{N_w+N_l}       & ~ \text{if }~ N_w> N_l\\
1  & ~ \text{if }~ N_w= N_l
\end{cases}
\end{align*}
Although using the exact test statistic appears to be the best choice when the sample size is small, it can be computationally intensive and is rarely used in practice. Furthermore, Fagerland et al. \cite{fagerland2013mcnemar, fagerland2014recommended} demonstrated that the ultra-conservatism of exact test produces unnecessary large p-values and is underpowered (see also Agresti and Coull \cite{agresti1998approximate}).  They recommend using, instead, the unconditional large-sample test statistic below.

\subsection{Test statistic based on a multinomial distribution} \label{sec:ls_test}
We can also leverage the multinomial distribution \eqref{eq:multi} to  test the difference between the two treatment groups, for the null hypothesis $H_0: \pi_w=\pi_l$ against the two-sided alternative $H_1: \pi_w\ne \pi_l.$ 

To do this, we consider the net benefit $D_w=p_w-p_l.$ Its mean and variance are
\begin{align*}
\mu(D_w)=E(D_w)&=E(p_w)-E(p_l)=\pi_w-\pi_l\\
\sigma^2(D_w)=Var(D_w)&=\text{Var}\, (p_w)+\text{Var}\, (p_l)-2\text{Cov}\, (p_w, p_l)\\
&=\displaystyle\frac{1}{N}\left[ \pi_w(1-\pi_w)+\pi_l(1-\pi_l)+2\pi_w\pi_l\right] =\displaystyle\frac{1}{N}\left[ \pi_w+\pi_l-(\pi_w-\pi_l)^2\right]. 
\end{align*} 
Under the null hypothesis $H_0$, these mean and variance are equal to 
\begin{align*} 
\mu_0(D_w)=0; ~~
\sigma_0^2(D_w)=\displaystyle\frac{1}{N}\left(\pi_w+\pi_l\right).
\end{align*} We can approximate the variance $\sigma_0^2(D_w)$ by $$\widehat \sigma_0^2(D_w)=\displaystyle\frac{1}{N}\left(p_w+p_l\right).$$
Using the normal approximation to the multinomial distribution, we derive a large-sample test for $H_0$   as 
\begin{align}\label{eq:test_multinomial}
Z_m&=\frac{D_w}{\sqrt{\widehat \sigma_0^2(D_w)}}=\frac{p_w-p_l}{\sqrt{(p_w+p_l)/N}}. 
\end{align} 
We show in Appendix \ref{appendix:test} that  the above test statistic $Z_m$ is exactly equivalent to the large-sample test statistic $Z$ derived  previously (see equation \eqref{eq:test_binom} in of Section \ref{sec:pocock_short}), using the normal approximation to the conditional binomial distribution \eqref{eq:binomial}, i.e.,  
\begin{align*}
Z_m&= \displaystyle\left(Q_w-\frac{1}{2}\right)\sqrt{4(N_w+N_l)}=Z.
\end{align*}

Finally, if we replace $\displaystyle p_w$ and $\displaystyle p_l$ by $\displaystyle p_w=\frac{N_w}{N}$ and   $\displaystyle p_l=\frac{N_l}{N}$,  it becomes 
\begin{align}\label{eq:zfinal}
Z=Z_m&=\frac{N_w-N_l}{\sqrt{(N_w+N_l)}}. 
\end{align} 
This indicates that the number of ties $N_t$ does not contribute directly to and has a negligible impact on the significance of the test statistic $Z$. Nevertheless, ties may have an impact on its effect size both in term of magnitude and precision. \cite{agresti2004effects} 


\subsection{Power and sample size calculation}\label{sec:ssize}
We can use the above unconditional large-sample test statistic $Z$ to estimate the sample size needed in planning a study in which two matched samples are compared with the goal to reach a specific power $1-\beta.$  From equation \eqref{eq:test_multinomial}, using the central limit theorem,  the power of the test statistic $Z$ can be approximated by 
\begin{align*}
1-\beta&=P(|Z|>z_{1-\alpha/2}| H_1)\approx \Phi \left(\frac{-\sigma_0^2(D_w) z_{1-\alpha/2}+|\pi_w-\pi_l|}{\sigma^2(D_w)}\right)
\end{align*} 
Using simple algebra, we have $$|\pi_w-\pi_l|=\displaystyle z_{1-\alpha/2}\sigma_0^2(p_w-p_l)+z_{1-\beta}\sigma^2(p_w-p_l).$$ Therefore, the total number of matched pairs needed to conduct a two-sided test with a significance level $\alpha$ and a power of detecting a significant difference of $1-\beta$ is 
\begin{align}\label{eq:power}
N&=\left[ \frac{z_{1-\alpha/2} \sqrt{\pi_w+\pi_l}+z_{1-\beta}\sqrt{\pi_w+\pi_l-(\pi_w-\pi_l)^2}}{(\pi_w-\pi_l)} \right]^2 
\end{align} 
Note that the same formula can be used whether we need to estimate the sample size for the net benefit or the win ratio. In each case, we only need to specify the anticipated proportion of untied ("win-or-lose") pairs  $\pi_{\text{wl}}=\pi_w+\pi_l$ as well as  the expected net benefit $D$ (or win ratio $R$) to estimate the sample size needed. 
For instance, since the proportion of  wins  $\pi_w=D+\pi_l$ for the  net benefit and $\pi_w=R\pi_l$ for the win ratio, we have  

\begin{align}
N&=\frac{\left[ z_{1-\alpha/2} \sqrt{\pi_{\text{wl}}}+z_{1-\beta}\sqrt{\pi_{\text{wl}}-D^2}\right]^2}{D^2}  
\end{align} 
for the net benefit and 
\begin{align}
N&=\frac{\left[ z_{1-\alpha/2}(R+1)+z_{1-\beta}\sqrt{(R+1)^2-(R-1)^2\pi_{\text{wl}}}\right]^2 }{(R-1)^2\pi_{\text{wl}}} 
\end{align}
for the win ratio.

\section{Confidence Interval Estimation}\label{sec:conf_int}
Although, we can leverage the binomial distribution \eqref{eq:binomial} to derive estimates of the $100(1-\alpha)\%$ exact confidence intervals for the net benefit and the win ratio,\cite{chan2003proving} such an attempt is pointless since exact confidence intervals tend to be  more conservative than desired or expected due to the discreteness of the binomial distribution.\cite{agresti2003dealing,newcombe1998interval,newcombe2012confidence,donner2012closed}  More generally, despite the dual relationship that exists between test statistics and the confidence intervals, in the sense that we can always derive a confidence interval from a given test statistic, it is not always guaranteed that we will obtain a genuine, good performing confidence interval. 

In this section, we estimate the confidence intervals for the risk ratio $R_w$ and the net benefit $D_w$ using either the normal approximation to the multinomial distribution of the random vector $(N_w, N_l, N_t)$  or inversion methods, where  the $100\%(1-\alpha)$ confidence interval for an estimand $\theta$ are the two solutions we obtain by solving for $\theta$  a score test equation of the form $|\widehat\theta-\theta|=z_{\alpha/2}\sigma_0(\widehat\theta)$, where $\widehat\theta$ is an estimator of $\theta$ and the standard error $\sigma_0(\widehat\theta)$ is evaluated at the null hypothesis.
\subsection{The matched net benefit}\label{sec:win_diff}
\subsubsection{The Wald method}
The Wald confidence interval is the most popular method of estimating the confidence intervals.   It assumes that the sampling distributions of $p_w$ and $p_l$ are asymptotically normal. 
For the matched net benefit $D_w=p_w-p_l$, using the distribution \eqref{eq:multi}, we have 
$\mu(D_w)=\pi_w-\pi_l$ and $$\displaystyle \text{Var}\,(D_w)=\frac{\pi_w+\pi_l+(\pi_w-\pi_l)^2}{N}.$$ We can obtain an estimate $\ \widehat\sigma^2({D_w})$ of the variance $\sigma^2(D_w)$  by replacing the probabilities $\pi_i$'s by their sample estimates, i.e.,
\begin{eqnarray}
\widehat\sigma^2({D_w})=\frac{ p_w+p_l-(p_w-p_l)^2}{N}=\frac{1}{N^2}\left[ N_w+N_l -\frac{(N_w-N_l)^2}{N}\right].
\end{eqnarray} 
Based on the central limit theorem, an approximate $100(1-\alpha)\%$ confidence interval for the true net benefit $D$ is given by 
\begin{eqnarray}\label{eq:nb}
(D_L, D_U)={D_w}\pm z_{\alpha/2}\widehat\sigma({D_w}).
\end{eqnarray}

Accordingly, since $D_w=p_w-p_l=(2Q_w-1)(p_w+p_l),$ using the confidence interval $(Q_L, Q_U)$ of $Q$ \eqref{eq:nus_binomial}, a large-sample $100(1-\alpha)\%$  confidence interval for the net benefit can also be determined   as 
$$\left(\frac{1}{N}(2Q_L-1)(N_w+N_l) ~,~ \frac{1}{N}(2Q_U-1)(N_w+N_l)\right) $$ 
which is equivalent to the confidence interval \eqref{eq:nb}.

The confidence interval issued from the above variance is commonly referred to as the Wald  confidence interval. 
For large sample sizes and probabilities $\ p_w$ and  $\ p_l$ away from  0 and 1, using the variance estimate $\widehat\sigma^2({D_w})$ can yield acceptable approximations of the $100(1-\alpha)\%$ confidence interval $(D_L, D_U)$ for the true net benefit $D$, as long as the underlying  assumption of normality for values of $p_w$ and $p_l$ holds. Samuels and Lu gave and investigated  a set of guidelines for deciding the situations when the Wald large-sample confidence interval provides good  estimations. \cite{samuels1992sample}

Although, the Wald confidence interval is the simplest and the most commonly-used confidence interval for the difference of proportions, several papers have highlighted its shortcomings. Indeed, unless the assumption of normality holds, Wald confidence intervals may yield inappropriate confidence limits that are out of the parameter space or confidence interval of zero width, even with moderate sample sizes.\cite{brown2001interval,brown2002confidence}  Moreover, for small sample sizes,  they may have poor and erratic coverage properties, especially when $\ p_w$ or  $\ p_l$ is at or near the boundaries 0 and 1.  \cite{newcombe1998interval,samuels1992sample,newcombe2012confidence,brown2001interval,brown2005confidence} In that case,  a different method such as the {\it square-and-add} Wilson score method  should be applied instead as it greatly improves performance, even under small sample size or when $\ p_w$ or  $\ p_l$ is at (or near)  0 or 1.\cite{newcombe2012confidence, bender2005number, connor2005proportions,newcombe1998interval,wilson1927probable}

\subsubsection{The MOVER method}\label{sec:wd_mover}
The square-and-add method combines two separate Wilson score confidence intervals for $p_w$ and  $p_l$ to form a confidence interval for their difference $p_w-p_l.$  The theoretical justification of the square-and-add Wilson method was given in a more general setting  by Donner and Zhou\cite{donner2012closed,zou2008construction} and summarized under the acronym {\it method of variance estimates recovery} (MOVER). Furthermore, Newcombe investigated and compared 11 methods for estimating the confidence interval for the difference of proportions and recommended the MOVER method.\cite{donner2012closed, newcombe1998two, newcombe2001estimating, newcombe2012confidence}

To construct the confidence interval of a function of two proportions, the MOVER uses separate, individual proportion confidence intervals and combine them into a single, hybrid confidence interval for the desired estimand. The idea behind the MOVER for the net benefit is simple; the method derives the variance of the difference by levering the variances of the individual proportions.
 Suppose, for each  $i=w, l$, there is a separate confidence interval $(L_i, U_i)$  which contains plausible values  of $\pi_i$. Our goal is to estimate the $100(1-\alpha)\%$  confidence interval $(D_L,D_U)$ of the net benefit $\text{NB}$ as a function of $L_i$ and $ U_i$,  $i=w, l$.
 
 By the central limit theorem, we can estimate the confidence interval  $(D_L,D_U)$ by $(D_L,D_U)=p_w-p_l\pm z_{\frac{\alpha}{2}}\{\widehat \text{Var}\,(p_w-p_l)\}^{1/2}$.
Unfortunately, as we already alluded to, such  confidence interval performs well only for large sample sizes or when the underlying distribution of the probabilities $p_w$ and  $p_l$ are truly normal. Nevertheless, thanks to the duality between a test statistic and the related confidence interval, we have $$\displaystyle \frac{\left[ \left(p_w-p_l \right) -\left(\pi_w-\pi_l\right)\right]^2 }{\text{Var}\,(p_w-p_l)}\leq z_{\frac{\alpha}{2}}^2~~\text{where}~~ \text{Var}\,(p_w-p_l)=\text{Var}\,(p_w)+\text{Var}\,(p_l)-2\text{Cov}\,(p_w, p_l).$$
Thus, we may view the  boundaries $D_L$ and $D_U$, respectively, as minimum and maximum of the values of the difference $\pi_w-\pi_l$ that satisfy 
i.e., \begin{eqnarray}\label{eq:cimover} 
	\displaystyle \frac{\left[ \left(p_w-p_l \right) -D_L\right]^2 }{\text{Var}\,(p_w)+\text{Var}\,(p_l)-2\text{Cov}\,(p_w, p_l)}= z_{\frac{\alpha}{2}}^2 ~~ ~~\text{and}~~ ~~\displaystyle \frac{\left[ \left(p_w-p_l \right) -D_U\right]^2 }{\text{Var}\,(p_w)+\text{Var}\,(p_l)-2\text{Cov}\,(p_w, p_l)}= z_{\frac{\alpha}{2}}^2, 
\end{eqnarray}
 which we can invert to derive the confidence interval $(D_L,D_U)$.

From the central limit theorem,  $( p_i-\pi_i)/\{\widehat \text{Var}\,(p_i)\}^{1/2}$ asymptotically follows the standard normal distribution as $N\longrightarrow\infty$. Hence, $(L_i, U_i) = p_i \pm z_{\frac{\alpha}{2}} \{\widehat \text{Var}\,(p_i)\}^{1/2}$, which leads to $\widehat \text{Var}\,(p_i)=z_{\frac{\alpha}{2}}^{-2}{(p_i-U_i)^2}$ and $\widehat \text{Var}\,(p_i)=z_{\frac{\alpha}{2}}^{-2}{(p_i-L_i)^2}$. Among the plausible values $\pi_i$ contained in $(L_i, U_i)$, those closest to the lower bound $D_L$ and the upper bound $D_U$ are, respectively, $L_w-U_l$ and $U_w-L_l$ , in the spirit of the score-type confidence interval.\cite{bartlett1953approximate} Therefore, to derive $D_L$, we estimate the variances of $p_w$ and $p_l$, respectively, at $L_w$ and $U_l$,  i.e., $\widehat \text{Var}\,(p_w)=z_{\frac{\alpha}{2}}^{-2}{(p_w-L_w)^2}$ and $\widehat \text{Var}\,(p_l)=z_{\frac{\alpha}{2}}^{-2}{(p_l-U_l)^2}$. Similarly, with   respectively $U_w$ and $L_l$, we have $\widehat \text{Var}\,(p_w)=z_{\frac{\alpha}{2}}^{-2}{(p_w-U_w)^2}$  and $\widehat \text{Var}\,(p_l)=z_{\frac{\alpha}{2}}^{-2}{(p_l-L_l)^2}$ to estimate the upper bound $D_U.$ 
Plugging these estimators back into the equation \eqref{eq:cimover}, we obtain
 $(D_L, D_U)$ in terms of the interval bounds $(L_i, U_i)$ of $\pi_i$  as 
\begin{align}\label{eq:mover_diff}
D_L=&p_w-p_l-\sqrt{(p_w-L_w)^2+(U_l-p_l)^2-2\widehat{ \rho}(p_w-L_w)(U_l-p_l)}\\
D_U=&p_w-p_l-\sqrt{(U_w-p_w)^2+(p_l-L_l)^2-2\widehat{\rho}(U_w-p_w)(p_l-L_l)}\nonumber
\end{align}
where $\displaystyle\widehat\rho=\displaystyle \widehat{ Corr}(p_w,p_l)$ is an estimator of the   correlation coefficient $\displaystyle \rho={ Corr}(\pi_w,\pi_l).$  
From the multinomial distribution \eqref{eq:multi}, we have  $\displaystyle\widehat\rho={-p_wp_l}/{\sqrt{p_w(1-p_w)p_l(1-p_l)}}$ if $p_w(1-p_w)p_l(1-p_l)\ne 0
$ and $\displaystyle\widehat \rho=0$ otherwise.

Studies by Agresti and Coull\cite{agresti1998approximate}, Newcombe\cite{newcombe2012confidence,newcombe1998two}, and Agresti and Caffo\cite{agresti2000simple} demonstrated, through extensive simulations, that the MOVER confidence interval performs much better than  the Wald confidence interval in terms of coverage probability and the mean length of confidence interval.

To estimate the confidence limits $(D_L, D_U),$ we need the individual confidence intervals $(L_i, U_i)$ for both $\pi_w$ and $\pi_l$ where $i=w, l$. We will not, however, use their corresponding standard Wald intervals as they are fraught with  limitations similar to those from the Wald two-sample proportion intervals, if not more.\cite{brown2001interval,agresti2005simple} In fact, the MOVER confidence interval reduces to the standard Wald confidence interval of the difference $p_w-p_l$ if we use Wald intervals limits of $\pi_w$ and $\pi_l$.\cite{donner2012closed}  Instead, we recommend using the confidence intervals for  $\pi_w$ and $\pi_l$ given by the following better-performing methods\cite{brown2001interval,brown2002confidence}

\begin{enumerate}
	\item the square-and-add Wilson intervals\cite{newcombe1998two}
	\begin{eqnarray}\label{eq:limW}
	(L_i, U_i)= \widetilde p_i  \pm \frac{0.5}{\widetilde{N}}z_{\frac{\alpha}{2}}\sqrt{z_{\frac{\alpha}{2}}^2+ 4N p_i(1-p_i)}
	\end{eqnarray}
	
	\item or the Agresti-Coull (also known as adjusted Wald) confidence intervals\cite{agresti1998approximate,brown2001interval}
	\begin{eqnarray}\label{eq:limA}
	(L_i, U_i)=\widetilde p_i\pm z_{\frac{\alpha}{2}}\displaystyle\sqrt{{\widetilde p_i\left(1-\widetilde p_i\right)}\big/{\widetilde N}}
	\end{eqnarray}
where $\displaystyle \widetilde p_i=\displaystyle \left(Np_i+0.5z_{\frac{\alpha}{2}}^2\right)\big/\widetilde{N} $ and $\widetilde N=  N+z_{\frac{\alpha}{2}}^2.$
\end{enumerate} 

\subsection{The matched win ratio}\label{sec:win_ratio}
In this section, we use this former relationship  to estimate the variance of win ratio $R_w$ using four different methods:  two delta methods based, respectively,  on $R_w$ and on $\log(R_w)$; the Fieller's theorem; and the MOVER methods using $Y=p_w-Rp_l$, where $\displaystyle R={\pi_w}/{\pi_l}$ is the true win ratio. 

\subsubsection{Wald methods} \label{sec:wald_ratio}
Assuming the asymptotic normality of the win ratio $  R_w=\displaystyle\frac{p_w}{p_l}$, we can apply the delta method to the function $\displaystyle f(x,y,z)=\frac{x}{y}$ to calculate the variance of $  R_w$ as $\displaystyle \sigma^2(R_w)= \text{Var}\,(R_w)=\frac{\pi_w(\pi_w+\pi_l)}{N\pi_l^3}$ (see Appendix \ref{appendix:var_wr}).  
We can replace the unknown parameters $ \pi_w$ and $ \pi_l$ by their corresponding unbiased and consistent estimators $\ p_w$ and $\ p_w$ to obtain an  estimate $\widehat\sigma^2({R_{w}}) $ of the variance
\begin{align}\label{eq:varWR}
\widehat\sigma^2({R_{w}}) = \frac{p_w(p_w+p_l)}{N\ p^3_l}.
\end{align}
Thus, the corresponding confidence interval is $\displaystyle R_w \pm  z_{\alpha/2}\widehat\sigma({R_w}).$

Note that $ R_w$ is the ratio of two sample proportions; the sampling distribution of  $ R_w$  can be skewed---especially when the sample size $N$ is not large enough---whereas the $\log(R_w)$ is usually relatively symmetric. Therefore, when $N$ is small such a variance estimation \eqref{eq:varWR} derived from the large-sample normal approximation of the sampling distribution of the win ratio may not be accurate enough to provide a good estimate of the $100(1-\alpha)\%$ confidence interval of $R$.

To avoid this problem and improve the normal approximation of the target statistic, we can use the logarithm transformation $\log(R_w)=\log(\pi_w)-\log(\pi_l).$ The logarithm transformation is widely used in the statistical literature to estimate confidence intervals for the risk ratio and the odds ratio,\cite{agresti2013categorical,katz1978obtaining,lachin2009biostatistical,rosner2015fundamentals}  assuming that $\log(R_w)$ is asymptotically normally distributed. By use of the delta method to the logarithm function, we find (see Appendix \ref{appendix:var_logwr})

\begin{align}\label{eq:waldWR}
\sigma^2( \log(R_w))= \text{Var}\left( \log(R_w)\right)  = \frac{1}{N}\left( \frac{1}{\pi_w} + \frac{1}{\pi_l} \right),
\end{align}
which is estimated as 
$
\displaystyle\widehat\sigma^2({\log(R_w)} )=  \frac{1}{N}\left( \frac{1}{p_w} + \frac{1}{p_l} \right)=\frac{1}{N_w} + \frac{1}{ N_l} .
$
Therefore, the $100(1-\alpha)\%$ confidence interval of  $R_{w}$ is  $\left(R_L, R_U\right) $  where $ R_L=R_w\exp\left( -z_{1-\frac{\alpha}{2}}\widehat\sigma_{\log}\right)~$ and $~ R_U=R_w\exp\left( z_{1-\frac{\alpha}{2}}\widehat\sigma_{\log}\right). $ 

There are three main limitations when using the above variance estimation \eqref{eq:waldWR}. First, the confidence interval cannot be computed when $p_w$ or $p_l$ is equal to 0. Second, since it is a large-sample approximation, this variance formula may be unreliable for small samples. Third, the method requires that the distributions of $\log(p_w)$ and $\log(p_l)$ to be asymptotically or approximately normally distributed, which is impossible since $p_w$ and $p_l$ are themselves asymptotically normally distributed. 
The method can lead to inflated variance estimations, which are inefficient since it provides confidence intervals that are unnecessary too wide.  Lui found that the logarithm transformation method could lead to confidence intervals that are much as 40 times wider than those from other competing methods.\cite{lui2007interval} 


\subsubsection{The Fieller's theorem method  }\label{sec:fieller}  
The variance  \eqref{eq:waldWR} depends on the normal approximation of $\log(R_w)$, which may still be skewed. To circumvent the issue of possible skewness in the sampling distribution of  $\log(R_w)$ in small sample size, we  use the Fieller's theorem\cite{fieller1954some, casella2002statistical} to better estimate the variance of the win ratio. For this purpose, we   consider the random variable $Y=p_w-Rp_l$.
The random variable $Y$ is the difference of two random variables, thus its sampling distribution is likely to be less skewed. Nevertheless, the Fieller's theorem method of confidence interval estimation assumes that probabilities $p_w$ and $p_l$ follow a joint bivariate normal distribution so that the difference $Y$ is normally distributed.

\begin{figure}[htbp]
	\centering
	\includegraphics[trim=25 30 40 35, clip, width=0.6\linewidth]{"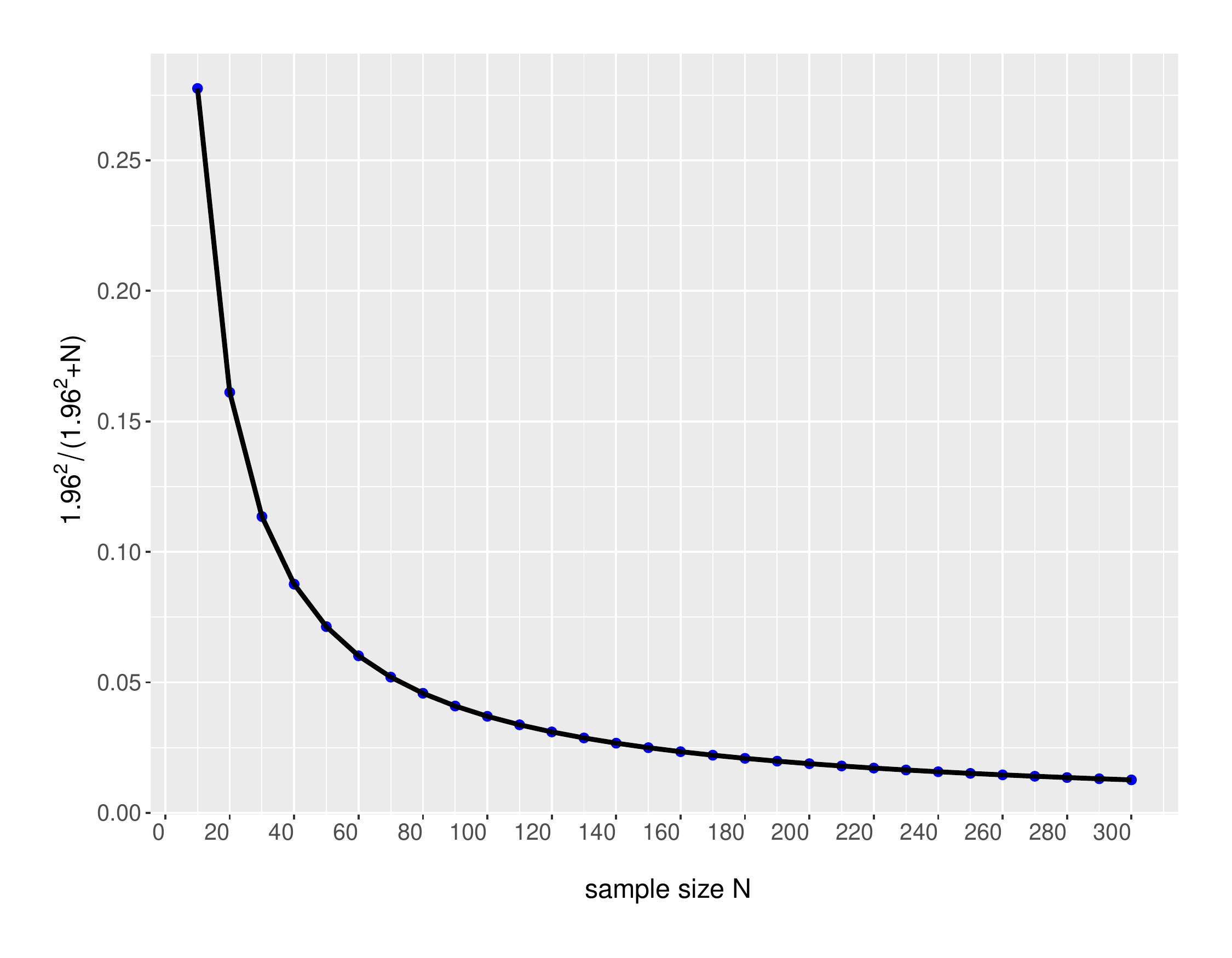"}
	\caption{Lower bound for $p_l$ to yield $A>0$, for  sample size N}
	\label{fig:filler_bound}
\end{figure}

Note that the mean $E(Y)= E(p_w)-RE(p_l)=\pi_w-\displaystyle \frac{\pi_w}{\pi_l} \pi_l=0.$ Using the delta method on the function $h(x,y,z)=x-Ry,$ we can derive the following asymptotic  variance of $Y,$ as shown in Appendix \ref{appendix:var_diffwr}:
\begin{align}\label{eq:varY}
\text{Var}\,(Y)  =\frac{1}{N}\left[  \pi_w(1-\pi_w)+2R\pi_w\pi_l+R^2\pi_l(1-\pi_l)\right]. 
\end{align}
The Fieller's method confidence interval for $R$ can then be determined by solving the following equation in $R,$
$$\displaystyle \frac{\sqrt{N}\left|Y\right|}{\sqrt{ p_w(1-p_w)+2Rp_wp_l+R^2p_l(1-p_l)}}\leq z_{\frac{\alpha}{2}}.$$ This is equivalent to solving the quadratic function $AR^2-2BR+C\leq 0$, where 
\begin{align*}
A &=  Np_l^2-z_{\frac{\alpha}{2}}^2p_l(1-p_l); ~
B=p_wp_l\left( N+z_{\frac{\alpha}{2}}^2\right);~
C= Np_w^2-z_{\frac{\alpha}{2}}^2p_w(1-p_w).
\end{align*}
Therefore, if $A>0$ and $B^2-AC>0,$ then the $100(1-\alpha)\%$ confidence interval of $R$ is given by $$ \left[ \max\{\left( B-\sqrt{B^2-AC}\right) /A~, ~0\}~,~ \left( B+\sqrt{B^2-AC}\right)/A\right].$$ 
However, if  $B^2-AC> 0$ but $A<0,$ the equation is concave which means the solution is a union of two disjoint open intervals $(-\infty, R_L)$ and $(R_U, \infty)$.  We  should also bear in mind that $A$ is negative whenever $p_l<z_{\frac{\alpha}{2}}^2/(z_{\frac{\alpha}{2}}^2 +N)$. This is fairly common when the sample size is small as shown in Figure \ref{fig:filler_bound} for $\alpha=0.05$ and illustrated in Section \ref{sec:application_udca}.
Furthermore, whenever $B^2-AC\leq 0$ (regardless of $A$) the two solutions to the  equation $AR^2-2BR+C= 0$ do not exist; the solution consists of the whole line $(-\infty, \infty)$.  Therefore, in either of these last two scenarios, the corresponding confidence for $R$ does not exist;  the Fieller's method is thus inapplicable.

The Fieller's theorem method  has also other important limitations. First, for some values of $p_w$ or $p_l,$ the confidence interval for the win ratio $R_w$ may not exist for a given data set if $B^2-AC\leq 0$. Second, the performance of the confidence interval determined via the Fieller's theorem relies crucially on the assumption that the joint distribution of  $p_w$ and $p_l$ is bivariate normal,\cite{zou2010generalization} which may not always be a good approximation, particularly when samples sizes are small. In fact when either $p_w$ or $p_l$ or both have large variances or if the sample size is small, the assumption of a bivariate normal distribution for $p_w$ and $p_l$ is not satisfied. Third, assuming that the condition  $B^2-AC> 0$ is met, it is still mathematically possible for the confidence interval, determined by solving the equation above, to contain negative values. In that case, the lower limit will be truncated at 0. The interval will contain only positive values if and only if $AC>0,$ that is, if only if both $p_w$ and $p_l$ are both significantly different from 0. That both $p_w$ and $p_l$ must be statistically significantly different from 0 is worrisome: if one or both have large variances or if the sample size is small, the condition will not be satisfy.

\subsubsection{The MOVER method  }\label{sec:mover_wr}
We consider the results from the MOVER method of section \ref{sec:wd_mover}---which is based on the difference of two random variables---to derive a confidence interval  $(R_L, R_U)$ for the win ratio $R$.\cite{donner2012closed,fagerland2014recommended, tang2012confidence2} 

Using equation \eqref{eq:mover_diff}, the confidence interval of  $Y=\pi_w-R\pi_l=0$ can be estimated as 
\begin{align}\label{eq:mover_ratio}
Y_L&=p_w-Rp_l-\sqrt{(p_w-L_w)^2+R^2(U_l-p_l)^2-2R\widehat{ \rho}(p_w-L_w)(U_l-p_l)}\\
Y_U&=p_w-Rp_l-\sqrt{(U_w-p_w)^2+R^2(p_l-L_l)^2-2R\widehat{\rho}(U_w-p_w)(p_l-L_l)}\nonumber
\end{align}
Hence, because $Y=0$, we set $Y_L=0$ and $Y_U=0,$ and solve for $R$ each time to obtain the limits $(R_L, R_U)$ of the $100(1-\alpha)\%$ confidence  interval of the true  win ratio  $R$ as    
\begin{eqnarray*}
	R_L&=&\frac{\left[ p_wp_l-\widehat{ \rho}(p_w-L_w)(U_l-p_l)\right]}{U_l(2p_l-U_l)} -\frac{\sqrt{\left[ p_wp_l-\widehat{ \rho}(p_w-L_w)(U_l-p_l)\right]^2-L_wU_l (2p_w-L_w)(2p_l-U_l)}}{U_l(2p_l-U_l)}\\
	R_U&=&\frac{\left[ p_wp_l-\widehat{ \rho}(U_w-p_w)(p_l-L_l)\right]}{L_l(2p_l-L_l)} +\frac{\sqrt{\left[ p_wp_l-\widehat{ \rho}(U_w-p_w)(p_l-L_l)\right]^2-U_wL_l (2p_w-U_w)(2p_l-L_l)}}{L_l(2p_l-L_l)}.\nonumber
\end{eqnarray*}
The $100(1-\alpha)\%$ confidence interval from the MOVER for ratios does not require that the distribution of $p_w$ and $p_l$ be symmetric. 
Note that unlike the Fieller's theorem method confidence interval estimation, the MOVER method does not require symmetry of the sampling distributions of $p_w$ and $p_l$ and always produces a bound-respecting confidence interval. 

To estimate the interval limits $(L_i, U_i)$ of the individual proportions $p_i$, $i=w, l$, we use the individual square-and-add Wilson intervals given in equation \eqref{eq:limW} or the Agresti-Coull intervals \eqref{eq:limA}, as we indicated in Section \ref{sec:win_diff}.

\section{Simulations}\label{sec:simulations}
In this section, we run sets of simulation studies to (1) evaluate the performance of the tests statistic of Section \ref{sec:tssize} and (2) compare the different confidence interval methods for the net benefit and the win ratio presented in Section \ref{sec:conf_int}, for  $\alpha = 0.05$.  In each scenario, we consider small $(N\leq 40)$ and moderate ($40<N<200$)  sample sizes, when necessary. To compare  the different methods, we   generated $10^5$ random samples of $(n_w,n_l,n_t)$ from a multinomial distribution $ M(N, \pi_w,\pi_l,\pi_t),$ under the null $H_0:\pi_w=\pi_l$ or the alternative hypothesis $H_1:\pi_w\neq \pi_l$, for given sample size $N$ and probability $\pi_w$ $,\pi_l$, or $\pi_t$ specified below.

\subsection{Type I error and power of the tests statistic}
We compare the performance of the test statistic $Z_P$  \eqref{eq:test_binomial} suggested by Pocock et al. \cite{pocock2011win}  and our proposed test statistic $Z$ given by \eqref{eq:zfinal},  in terms of type I error and power calculation.  
To estimate the type I error, we fixed  $N=30, 40, 50$ as well as $N=100$ and 200. We assumed that $\pi_w=\pi_l$ and varied their values between 0.1 and 0.5 by an increment of 0.1.

\begin{figure}
	\centering
	
	\centering
	\includegraphics[width=0.9\linewidth]{"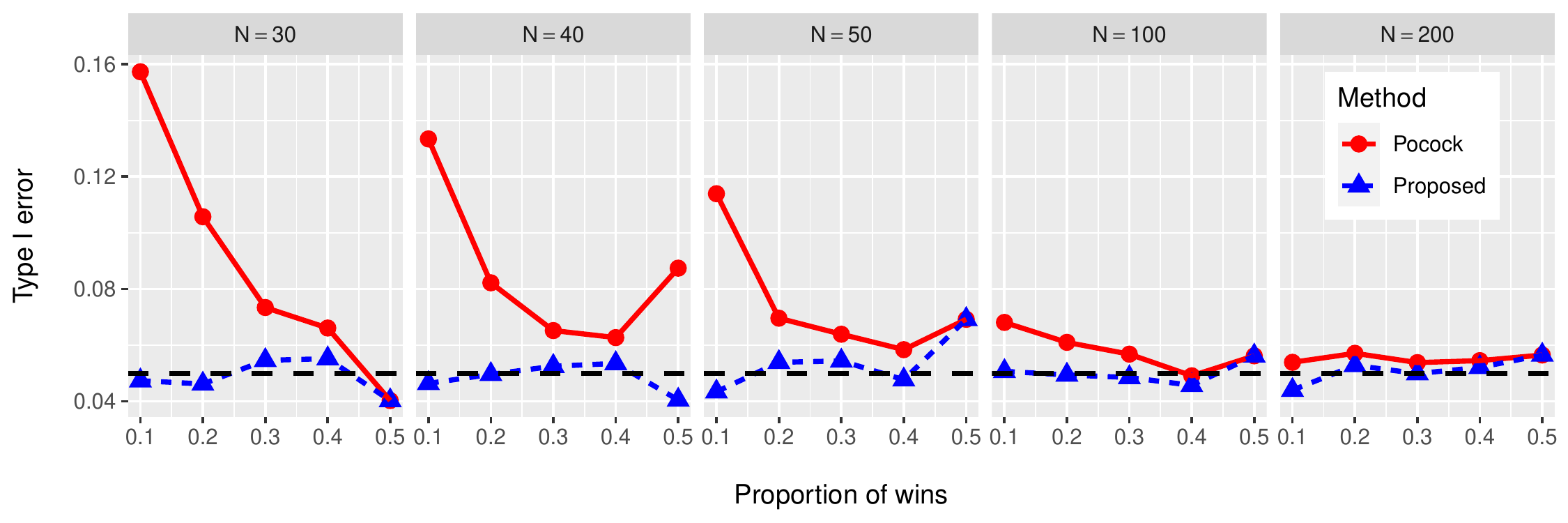"}
	\caption{Estimated type I error of the Pocock et al. and our proposed tests statistic.}
	\label{fig:typeIerror}
\end{figure}

We present the results in Figure \ref{fig:typeIerror} (and in Table \ref{tab:typeIerror} in the Appendix); the  black horizontal dashed line represents the true type I error  $\alpha = 0.05$. As expected, the test statistic $Z_P$   provides an adequate type I error only when the sample size is large ($N\geq 200$). The mean type I error estimate is between 0.05 and 0.06. However, when the sample size is smaller than or equal to 100, the test statistic $Z_P$ over-estimates the Type I error across for different values of the proportions of wins. The estimate gets worse as the sample size decreases (for fixed $p_w$) or the proportion of wins decreases (for a fixed sample size). On the other hand,  the proposed test statistic $Z$  controls the type I error fairly well, regardless of the sample size or the proportion of wins.

\begin{figure}
	\centering
	\includegraphics[width=0.9\linewidth]{"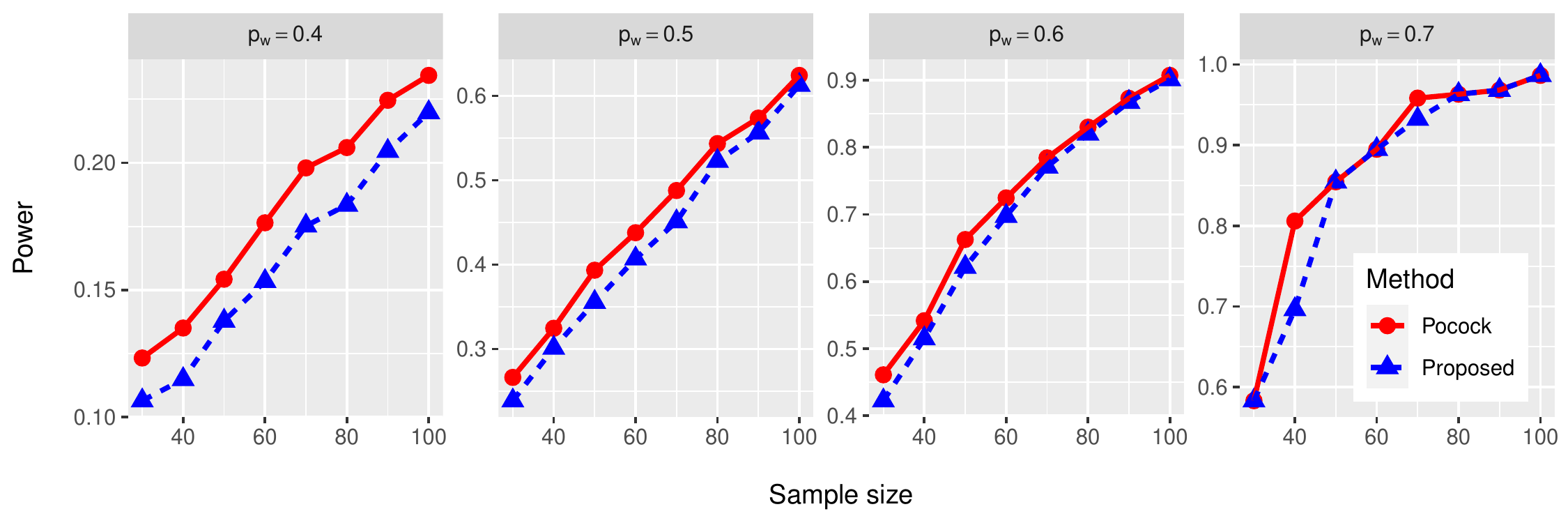"}
	\caption{Empirical power of the Pocock et al. and our proposed tests statistic, by proportion of wins $p_w$ and sample size.}
	\label{fig:power}
\end{figure}
For power calculation, we chose   $\pi_l=0.3$ and considered three different values of $\pi_w$, i.e., $\pi_w=0.4, 0.6$ and 0.6.  We estimated the power for $N=30$ to $N=100$, by an increment of 10. 
The results  are given in Figure \ref{fig:power} (and in Table \ref{tab:power}  in the Appendix). For any proportion of the proportion of wins and any sample size, the test statistic $Z_P$ from Pocock et al.\cite{pocock2011win} has always a higher power than our proposed test statistic $Z$. This pattern is not surprising. Rather similar to the type I error, the test statistic from  Pocock et al.\cite{pocock2011win} is over-optimistic and thus tend to reject the null hypothesis more than it should. Nevertheless, as expected, the gap between the two curves narrows and finally closes as the sample size or the proportion of wins increases.

Overall, our proposed test statistic $Z$  controls the type I error and provide a better (and honest) power estimation compared to the test statistic $Z_P$  from Pocock et al. \cite{pocock2011win}

\subsection{Confidence intervals for the net benefit and the win ratio} 
  
We also conducted simulation studies to evaluate the small and moderate sample performance of each confidence interval presented in Section \ref{sec:conf_int} for the win ratio and the net benefit. We used two main criteria to assess the performance of confidence interval estimations:  the (empirical) coverage probability (CP)---the probability that the confidence interval contains the true value---and average confidence width (CW), i.e., width of the confidence interval. Although the target coverage probability cannot be achieved exactly since $N_w, N_l,$ and $N_t$ are discrete random variables, a confidence interval estimation is considered good if its coverage probability approximates the nominal 95\% coverage level. Overall, we expect the mean coverage to be slightly conservative (i.e., at least equal to 95\%) and if two methods provide an adequate coverage,   the one with the smaller confidence  width is preferred. \cite{newcombe2012confidence} 

To investigate the performance of the confidence intervals of the net benefit (resp. the win ratio), we   set the true net benefit to NB = 0.25, 0.375, and 0.5 (resp. the true win ratio WR = 1, 1.5 and 2). Then, in each case, we varied the proportion of ties $\pi_t$ from 0.1 to 0.5 by an increment of 0.1 and  generated the triplet $(N_{w}, N_{l}, N_{t}) \sim M(N, \pi_{w},\pi_{l},\pi_{t})$. 
Varying the proportion of ties allowed us to explore different values of the proportions of wins and losses. From   the $10^5$ data replicates, we  calculated the probabilities $(p_{w}, p_{l}, p_{t})$ and determined both the coverage probability and the width of different proposed confidence intervals for small ($N = 30$  and 50) and moderate ($N=100$) sample sizes. 

\subsubsection{Confidence intervals for the net benefit} 

\begin{figure}
	\centering
	\includegraphics[trim=3 5 3 7, clip, width=0.9\linewidth]{"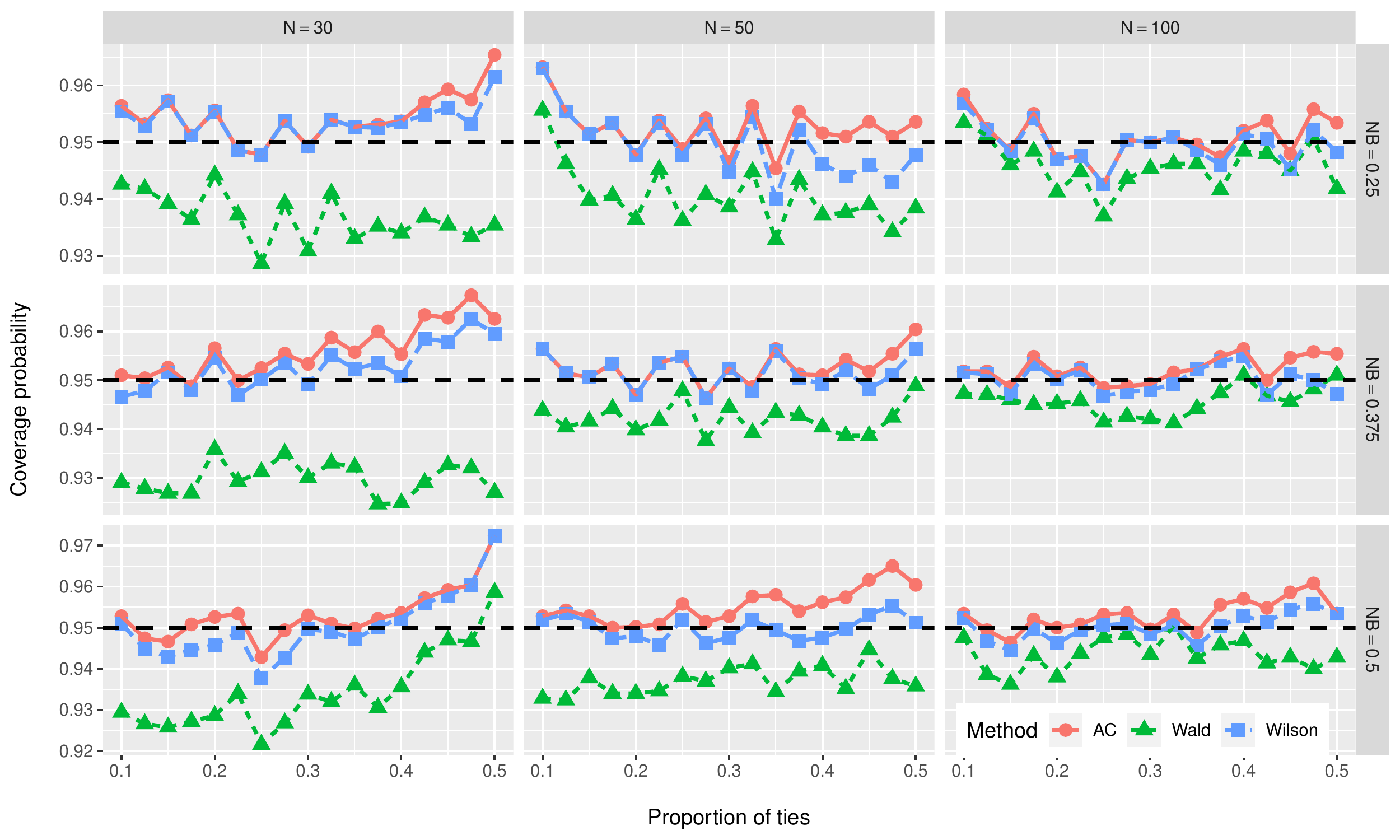"}
	\caption{Estimated coverage probability of the 95\% confidence interval  of the net benefit.}
	\label{fig:coverage_nb}
\end{figure}

\begin{figure}
	\centering
	\includegraphics[trim=3 5 3 7, clip, width=0.9 \linewidth]{"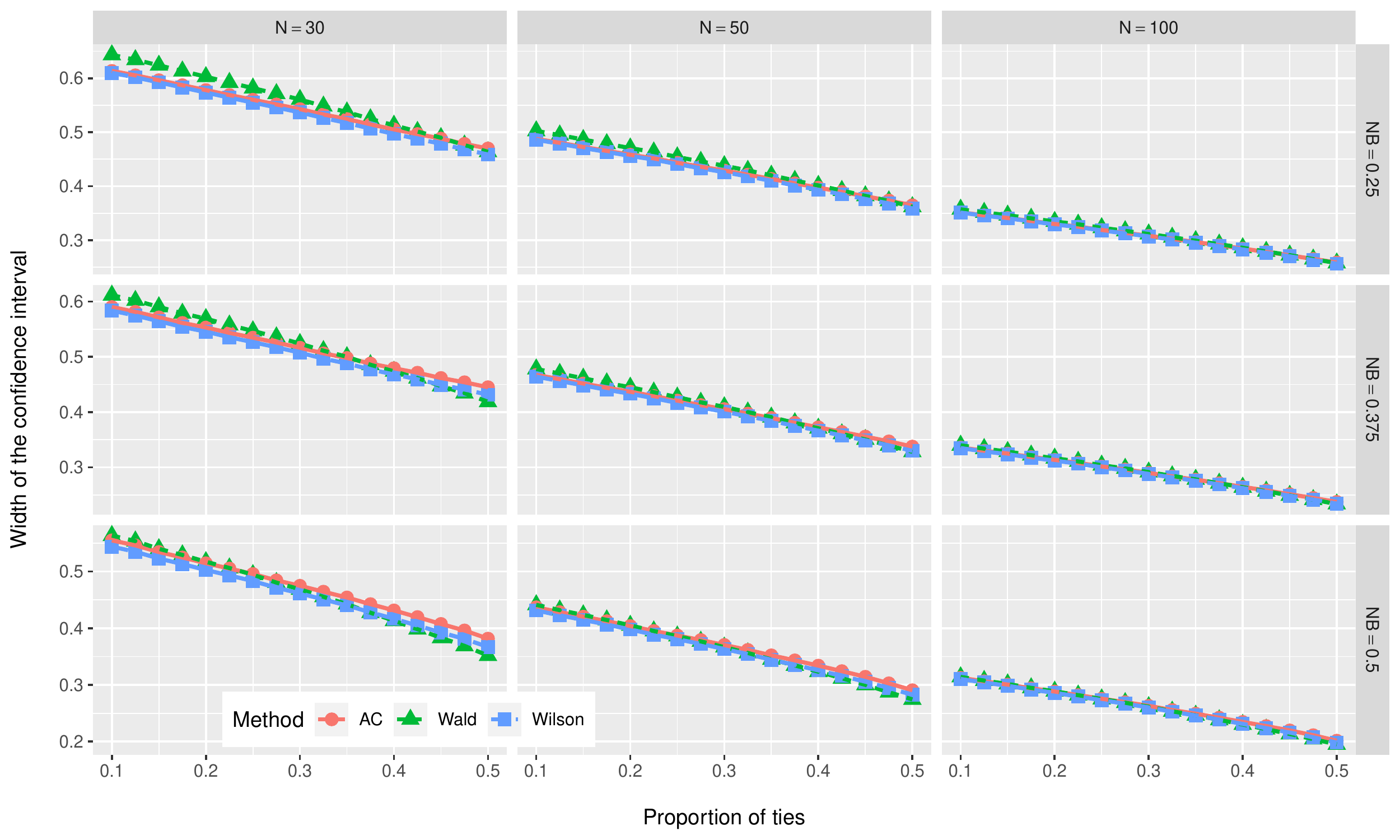"}
	\caption{Average width of the 95\% confidence interval of the net benefit.}
	\label{fig:width_nb}
\end{figure}

For the net benefit, we  generated $(N_{w}, N_{l}, N_{t}) \sim M(N, \pi_{w},\pi_{l},\pi_{t})$ with $\pi_{w}=(1+\text{NB}-\pi_{t})/2$ and $\pi_{l} =\pi_{w}- \text{NB}$ and compared the three methods to estimate the confidence interval presented in Section \ref{sec:win_diff}. The results are presented in Figures \ref{fig:coverage_nb} and \ref{fig:width_nb}, as well as in Tables \ref{tab:coverage_nb} and \ref{tab:width_nb} for the coverage and width of the 95\% confidence intervals of the net benefit  based on Wald method, the MOVER plus Agresti-Coull (AC), and the MOVER plus Wilson (Wilson).
   
As shown by Figure \ref{fig:coverage_nb}, all the CP curves have  broken, seesaw-like patterns with the Wald coverage curve harboring the steepest dips. When $N=30,$ the Wald method performs poorly with most coverage between 0.93 and 0.94, except for few cases such as when NB $=0.5$ and $\pi_t>0.4$. The coverage improves as the sample size increases, but remain mostly  below 0.95. The coverage of the AC method is quite close to the nominal level of 95\% except in few cases and always slightly above the Wilson coverage curves. In term of the width of the confidence intervals (see Fig. \ref{fig:width_nb}), the Wilson method has the lowest width followed by the Agresti-Coull and then the Wald method. The difference in the confidence width is  more pronounced when the sample size is small. Nevertheless, the gaps between the tree confidence width curves narrow significantly as the sample size increases and close when $N\geq 200$.

Overall, the Wald confidence interval is wider and anti-conservative, rarely achieving the nominal 95\%  coverage level. In general, the coverage of the confidence interval for both MOVER methods are a little above the nominal 95\% level, with  the Agresti-Coull coverage curve lying above than  the Wilson coverage curve, in most cases.  Therefore, taking both the coverage and the coverage into account, the Wilson method has the best confidence interval estimation, followed by the Agresti-Coull method and then the Wald method. 


\subsubsection{Confidence intervals for the win ratio}  
\begin{figure}
	\centering
	
	\centering
	\includegraphics[trim=4 6 7 7, clip, width=0.9\linewidth]{"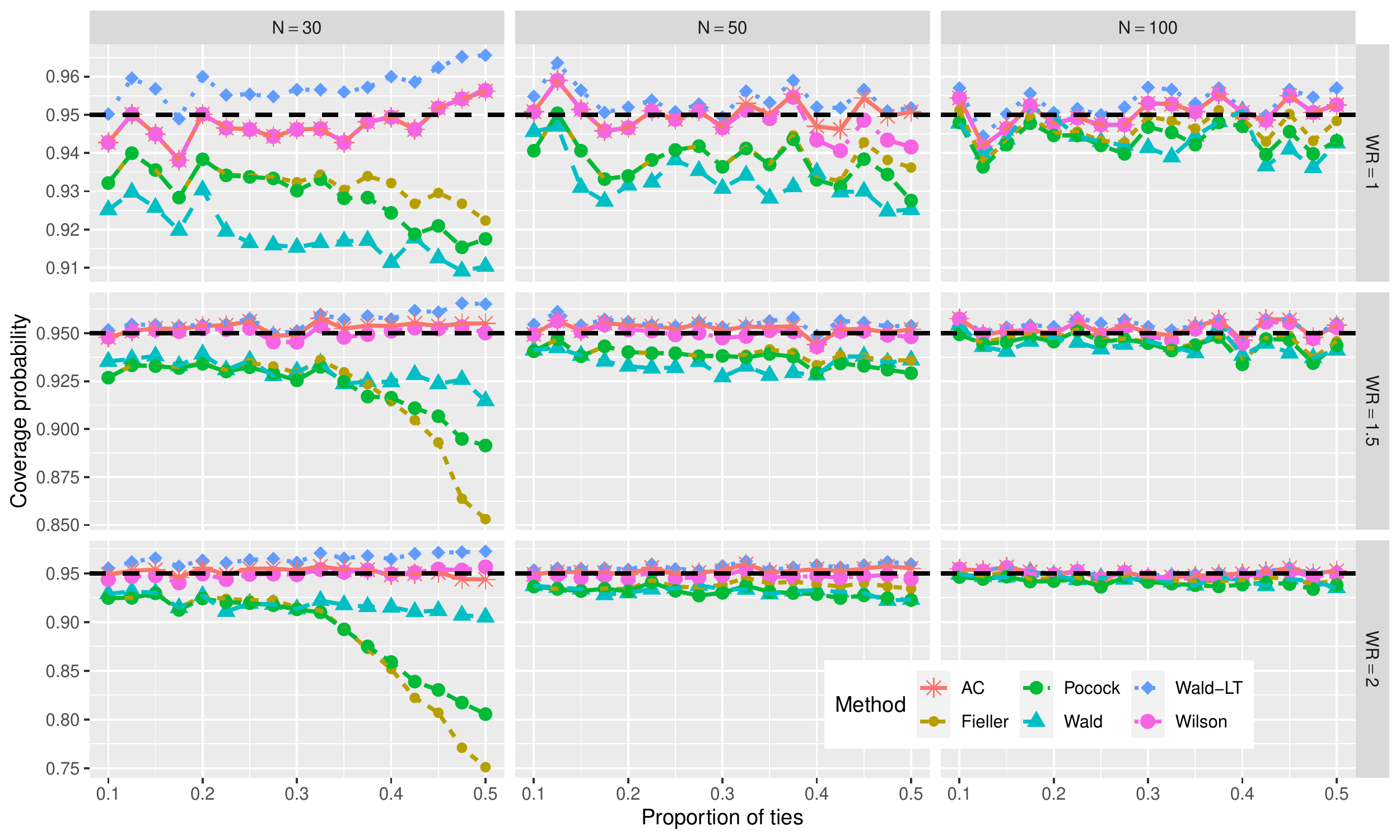"}
	\caption{Estimated coverage probability of the 95\% confidence interval of the win ratio.}
	\label{fig:coverage_wr}
\end{figure}

\begin{figure}
	\centering
	\includegraphics[trim=3 5 7 7, clip, width=0.9\linewidth]{"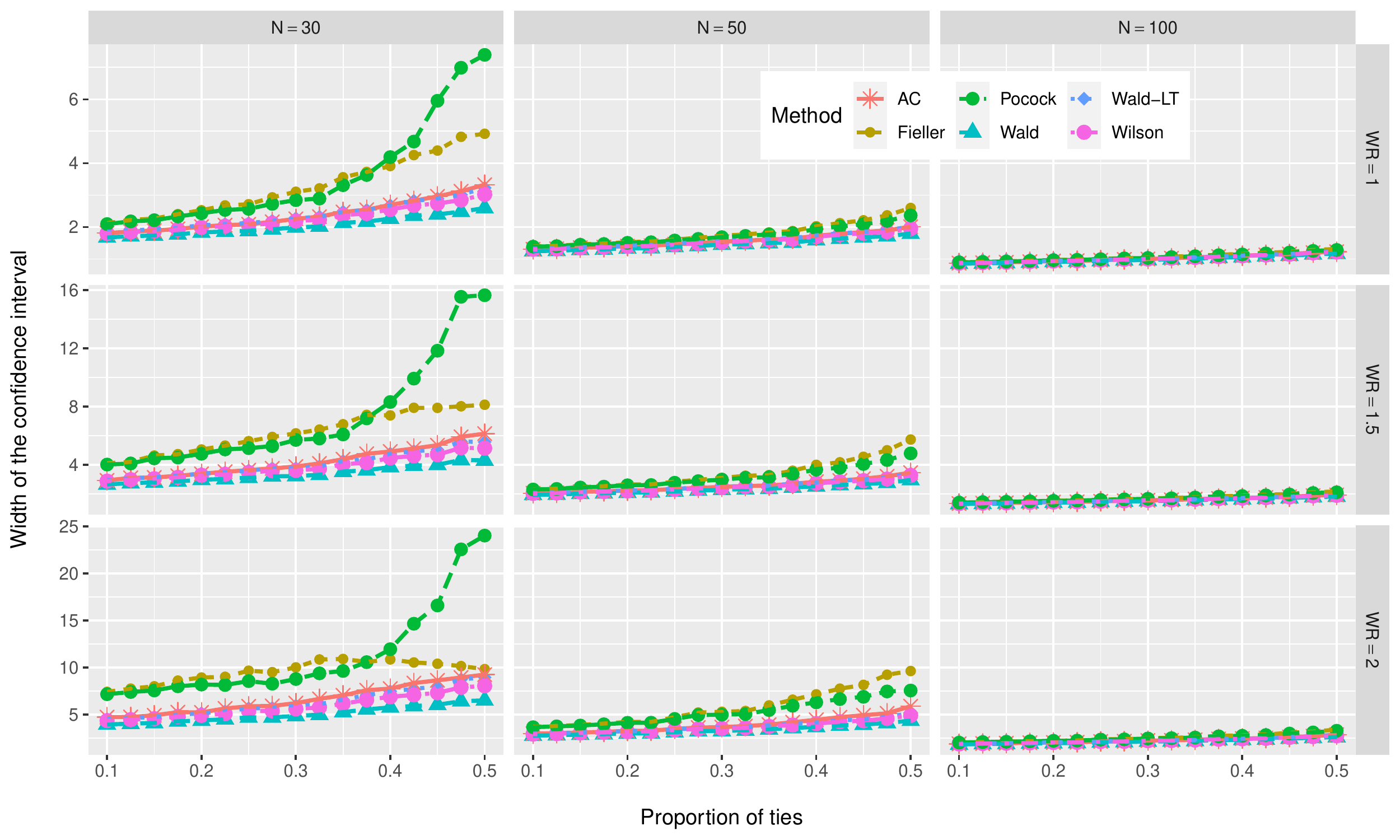"}
	\caption{Average width of the 95\% confidence interval of the win ratio.}
	\label{fig:width_wr}
\end{figure}

To compare the 6 confidence intervals of the win ratio, we generated $(N_{w}, N_{l}, N_{t}) \sim {\rm \mbox{Multinomial}}(N, \pi_{w},\pi_{l},\pi_{t})$ for  $\pi_{l}=(1-\pi_{t})/(1+\text{WR})$ and $\pi_{w} =\text{WR}\pi_{w}$. The results for the win ratio are shown in  Figures \ref{fig:coverage_wr} and \ref{fig:width_wr} and in Tables \ref{tab:coverage_wr} and \ref{tab:width_wr} for the Pocock et al.\cite{pocock2011win} method \eqref{eq:wr_binomial} as well as the 5 different  confidence interval methods of Section \ref{sec:win_ratio}: the (large-sample) Wald methods (Wald and Wald-LT), the Fieller's Theorem (Fieller) method , the MOVER plus Agresti-Coull (AC) and the MOVER plus the Wilson  (Wilson). 

Figure \ref{fig:coverage_wr} shows that the Wald,  Pocock, and Fieller's methods perform poorly, especially when the sample size is small ($N=30$). When  $N=30$, the three methods have a very poor coverage---at most 0.94 and below 0.93 when  $\pi_t\geq  0.25$. The lowest coverage  is reached  at $\pi_t= 0.5$; it is 0.75 for the Fieller's method, 0.81 for Pocock et al.'s method,  and 0.91 for the Wald method.  
Although the coverage under these methods improves as the sample size increases to $N=50$, it remains below 0.95. At $N=100,$ the coverage of the three is acceptably close to 0.95, particularly when WR $=2,$ but still below 0.95.

Both MOVER (via Agresti-Coull or Wilson) have generally very similar  coverage, mostly around the nominal level of 95\% regardless of the sample size or the magnitude of the treatment effect. The Wald confidence interval based on a logarithm transformation (Wald-LT) is conservative with the coverage at or (mostly) above 0.95.

Variations in average confidence widths were markedly present in lower sample size (see Fig. \ref{fig:width_wr}). When $N=30,$ the Wald method has the lowest average width followed by the MOVER (via Agresti-Coull and Wilson) and the  Wald-LT method, and then the Fieller's and Pocock et al. methods  tandem. The confidence width of the MOVER (via Agresti-Coull and Wilson) and Wald-LT method are, in general, similar throughout. In most cases, confidence widths of Pocock et al. method are below or equal to those from the Fieller's method. The only exceptions are when  $N=30$ and $\pi_t> 0.35$. From that specific point on, the curve of confidence width from the Pocock method increases steadily and the gap between the two curves widens as  $\pi_t$  increases. 
At $\pi_t= 0.5$, the gap is equal to 4.37 when WR $=1$, 7.52 when WR $=1.5$, and 14.19 when WR $=2$.

Overall, the MOVER (via Agresti-Coull and Wilson) and the Wald  method based on a logarithm transformation (Wald-LT) provide better confidence interval estimations compare to Pocock, Fieller's, and the large-sample Wald methods.

\section{Data applications}\label{sec:application}
We now compare different confidence interval estimation methods using  classic examples. The first two examples are  from Pocock et al.'s paper \cite{pocock2011win} and the last two are preeminent examples form the survival literature. The objective is to demonstrate that for (fairly) moderate to large sample size all the methods provide similar confidence intervals, but the results differ drastically when the sample size is small or the proportion of wins (or losses) is small.
\subsection{Data example 1: The EMPHASIS-HF trial} \label{sec:application_emphasis}
For our first example, we consider the EMPHASIS-HF clinical trial where patients aged 55 years or older  with NYHA functional class II symptoms and an ejection fraction of less than 35\% were recruited at 278 centers in 29 countries and followed for a median follow-up time of 21 months. Of the 2737 enrolled in the trial, 1364 were assigned to eplerenone and 1373 to placebo.\cite{zannad2011eplerenone,pocock2011win} To estimate the win ratio, Pocock et al. matched patients by their risk scores, which were derived  using a Cox proportional hazards model on a composite endpoint of death or hospitalization for heart failure.\cite{pocock2011win} There were 1364 pairs of patients with 249 wins in the treatment group, 151 wins in the placebo group and 964 ties. 

We have, $Z= 4.9$ with  a p-value of $4.8\times 10^{-7}.$ The net benefit NB $=0.07$ and win ratio WR $=1.65$ as well as their corresponding confidence intervals and widths are given in Table \ref{table:emphasis}. As expected for such a large sample size and $p_w=0.18$, all the methods yield similar confidence intervals for the net  benefit and win ratio.
\begin{table}
	\caption{The EMPHASIS-HF trial and CHARM Program}\label{table:emphasis}
	\begin{center}
		\begin{threeparttable}
			\begin{tabular}{cccccccccccccccccccccccccccc}
				\toprule				
				& &\multicolumn{3}{c}{The EMPHASIS-HF trial}&&\multicolumn{3}{c}{ The CHARM Program}\\
				\cmidrule(lr){3-5} \cmidrule(lr){7-9}		
				{ Estimand} &\multicolumn{1}{c}{ Method}& {Estimate}&{ 95\% CI}&{ 95\% CW}&& {Estimate}&{ 95\% CI}&{ 95\% CW}\\
				\cmidrule(lr){1-9}
				\multirow{3}{*}{Net Benefit}&Wald & 0.07 & (0.04, 0.10) & 0.06 & & 0.08 & (0.03, 0.12) & 0.09\\ 
				&MOVER A-C & 0.07 & (0.04, 0.10)&0.06 &&  0.08& (0.03, 0.12) & 0.09\\ 
				&MOVER Wilson & 0.07 & (0.04, 0.10)& 0.06&& 0.08 & (0.03, 0.12)  & 0.09\\ 
				\midrule
				\multirow{5}{*}{ Win Ratio}& Pocock & 1.65 & (1.35, 2.03) & 0.68 && 1.30 & (1.13, 1.50)  & 0.37\\
				&Wald&  1.65 & (1.32, 1.98) &0.66  &&  1.30 & (1.11, 1.49)  & 0.38\\ 
				&Wald-LT & 1.65 & (1.35, 2.02) & 0.67 && 1.30 & (1.12, 1.50)  & 0.38\\ 
				&Fieller's & 1.65 & (1.35, 2.03) &0.68  && 1.30 & (1.13, 1.50)  & 0.37\\  
				&MOVER A-C& 1.65 & (1.35, 2.02) & 0.67 && 1.30 & (1.12, 1.50) & 0.38\\  
				&MOVER Wilson & 1.65 & (1.35, 2.02)& 0.67 && 1.30 & (1.12, 1.50) & 0.38\\
				\bottomrule
			\end{tabular}
			\begin{tablenotes}\footnotesize
				
				\item 
				CI (resp. CW): Confidence interval (resp. width); Wald LT: Wald CI via log transform; MOVER A-C (resp. Wilson): MOVER CI via Agresti-Coull (resp. Wilson) individual proportion CI estimation approach
			\end{tablenotes}
		\end{threeparttable}
	\end{center}
\end{table}

\subsection{Data example 2: The CHARM Program} \label{sec:application_charm}
The second example, based on the CHARM Program,  consider a study of candesartan in patients with chronic heart failure who were randomoized either to candesartan or placebo and followed up
for at least 2 years. There were three types of patients whose  left-ventricular
ejection fraction (LVEF) was less than or equal to 40\%: those (1) who were not receiving
angiotensin-converting-enzyme (ACE) inhibitors because of previous
intolerance (CHARM Alternative) or (2) who were currently receiving ACE
inhibitors (CHARM Added), and (3) those with LVEF higher
than 40\% (CHARM Preserved). To illustrate the proposed methods, we considered only the CHARM Added type of patients and compared canderstan against placebo. 

Overall, 2548 CHARM-Added-type patients  were
randomly assigned to either candesartan (1276 patients) or placebo.\cite{pfeffer2003effects,pocock2011win} Similar to the EMPHASIS clinical trial, patients in canderstan and placebo were matched based on a risk score derived from a Cox proportional hazards model. There were 1272 pairs of patients with 421 wins in the candesartan group,  324  in the placebo group, and 527 ties. The proportion of wins is $p_w=0.33;$ the test statistic $Z=3.55$ with a p-value of $4\times 10^{-4}$. Using the different methods we presented in the previous sections, the net benefit NB $=0.08$ and win ratio WR $=1.30$  with their corresponding confidence intervals and widths given in Table \ref{table:emphasis}. The results are fairly similar  for the different methods we considered.

\subsection{Data example 3: UDCA Study in patients with primary biliary cirrhosis.} \label{sec:application_udca}
Our last example uses data from a Mayo Clinic double-blind randomized study on the effect of Ursodeoxycholic acid (UDCA) in the treatment of primary biliary cirrhosis. The data ({\it udca2}  from the R package {\it survival}) consist of 180 patients of whom  89 were assigned to the UDCA treatment group. To evaluate the incidence and (time to) treatment failure, the investigators considered 8 primary time-to-event endpoints: death, transplant, histological progression, development of varices, development of ascites, development of encephalopathy, worsening of symptoms, and doubling of bilirium. The number of events by treatment groups for each outcomes are shown in Table \ref{table:udca_events}.   A decision was made later not to use doubling of bilirium  and voluntary withdrawal from the study as endpoints due to, respectively, lab variation and possible bias. \cite{lindor1994ursodeoxycholic} 
                  
\begin{table}[h]
	\caption{ Number of events per outcome and by treatment group in the UDCA Study}\label{table:udca_events}
	\begin{center}
		\begin{threeparttable}
			\begin{tabular}{rccccccccccccccccccccccccccc}
				\toprule	
								{ } &\multicolumn{2}{c}{ Treatment}\\\cmidrule(lr){2-3}				
				{ Enpoint} &\multicolumn{1}{c}{Placebo}&{ UDCA}\\
				\cmidrule(lr){1-3}
				 Death & 10 & 6  \\ 
				Transplant & 6 & 6 \\ 
				histological progression & 12 & 8 \\ 
				
				development of varices&  17 & 8\\ 
				development of ascites & 5 & 1  \\ 
				development of encephalopathy & 1 & 3\\  
				worsening of symptoms& 9 & 7\\  
				doubling of bilirium & 15 & 2\\
				\bottomrule
			\end{tabular}
%
		\end{threeparttable}
	\end{center}
\end{table}

 For illustrative purposes, we consider 3 sets of hierarchical composite endpoints: (1) death alone, (2) death and transplant, and (3) all the 7 endpoints  (without {doubling of bilirium}), in the specific order aforementioned. Patients in the UDCA group are matched to those in the placebo by the their risk score, leading to a total of 84 matched pairs. The results for the confidence intervals of the net benefit and the win ratio are reported in Table \ref{table:udca_results}.

 Using death alone as outcome,   we obtain 10 wins, 3 losses, and 71 ties in the UDCA treatment group,. The proposed test statistic $Z$ yield a p-value of 0.052 while for the Pocock et al. test statistic $Z_P$, the p-value is 0.021. The confidence intervals for the net benefit are fairly similar.  
 
 While the main conclusion is that there is no difference in mortality between the two treatment groups, Pocock et al. test statistic, lead to the opposite conclusion with a surprisingly, extremely wide confidence interval of the win ratio $(1.17,~575.59).$ The  upper bound of this confidence interval is extreme because $p_w=0.12$ and $Q_w=0.77$, which leads to $Q_L\approx 1$ (the upper limit  of the confidence interval of $Q_w$). Likewise, the Fieller's method leads to an unrealistic confidence interval, a union of two disjoint open intervals $(-\infty, -30.71)\cup (1.02, +\infty)$, because $A=  -0.03$ while $B = 0.37$ and $C = 0.79$ (where A, B, and C are defined as in Section \ref{sec:fieller}). Furthermore, the Wald method shows a boundary anomaly since the lower bound of the win ratio is negative, even though the win ratio is always non-negative. Although we may decide to truncate the lower limit and set it to be 0, such a practice is pointless in regard of the requirement of (asymptotic) normal distribution imposed by the  Wald method, which is likely not satisfied in this data set. More importantly, such an ad hoc truncation is not needed since there are other alternative methods that provide adequate and sensible confidence intervals for the win ratio as shown here. The Wald-LT and two MOVER methods provide reasonable confidence intervals. Of these three methods, the MOVER Wilson has the smallest width (10.39), followed by the Wald-LT (11.19) and the MOVER AC (15.9).
 
  \begin{table}[h]
 	\caption{The UDCA Study in patients with primary biliary cirrhosis}\label{table:udca_results}
 	\begin{center}
 		\begin{threeparttable}
 			\begin{tabular}{cccccccccccccccccccccccccccc}
 				\toprule
 				&	{ } &\multicolumn{6}{c}{Composite outcome}\\	\cmidrule(lr){3-8}	
 				&	{ } &\multicolumn{2}{c}{Death} &\multicolumn{2}{c}{Death and transplant}&\multicolumn{2}{c}{All 7 endpoints}\\	\cmidrule(lr){3-4}	\cmidrule(lr){5-6}\cmidrule(lr){7-8}			
 				{ Estimand} &\multicolumn{1}{c}{ Method}& {Estimate}&{ 95\% CI}& {Estimate}&{ 95\% CI}& {Estimate}&{ 95\% CI}\\
 				\cmidrule(lr){1-8}
 				\multirow{3}{*}{Net Benefit}
 				&Wald & 0.08 & (0.001, 0.16) & 0.10 & $(-0.01, 0.20)$ & 0.24 & (0.08, 0.40) \\ 
 				&MOVER A-C & 0.08 & $(-0.007, 0.18)$ & 0.10 & $(-0.01, 0.20)$& 0.24 & (0.07, 0.39)\\ 
 				&MOVER Wilson & 0.08 & $(-0.002, 0.17)$  & 0.10 & $(-0.01, 0.20)$& 0.24 & (0.07, 0.39)\\ 
 				\midrule
 				\multirow{5}{*}{ Win Ratio}
 				& Pocock & 3.33 & (1.17, 575.59) & 2.33 & (1.00, 9.08)  & 2.25 & (1.31, 9.08)\\
 				&Wald&  3.33 & $(-0.97, 7.63)$& 2.33 & (0.10, 4.56)& 2.25 & (0.92, 3.58)\\ 
 				&Wald-LT & 3.33 & (0.92, 12.11) & 2.33 & (0.90, 6.07) & 2.25 & (1.25, 4.05)\\ 
 				&Fieller's & 3.33 & $\mathcal{I}^*$& 2.33 & (0.93, 11.10)& 2.25 & (1.30, 4.54)\\  
 				&MOVER A-C& 3.33 & (0.92, 16.82)& 2.33 & (0.90, 6.41)& 2.25 & (1.26, 4.07)\\  
 				&MOVER Wilson & 3.33 & (0.97, 11.33) &  2.33 & (0.92, 5.91)& 2.25 & (1.26, 4.04)\\
 				\bottomrule
 			\end{tabular}
 			\begin{tablenotes}\footnotesize

				\item	
				  $^*\mathcal{I} =(-\infty, -30.71)\cup (1.02, +\infty)$
 			\end{tablenotes}
 		\end{threeparttable}
 	\end{center}
 \end{table}

 With the composite outcome of death and transplant, there are 14 wins, 6 losses, and 64 ties for the UDCA group. The p-values are 0.05 and 0.07 for, respectively, for the Pocock et al. and our proposed test statistic. Again, all the methods yield the same confidence interval $(-0.01, 0.20)$ for the net benefit.  
 For the win ratio, all the methods, but the Wald method, have similar lower bounds. The  Fieller's Theorem and  Pocock et al. methods have wide confidence intervals, with respective widths equal to equal to 10.17 and 8.08. The MOVER  Wilson, the Wald-LT method, and MOVER AC  have smaller width, respectively 4.99,  5.17, and 5.51. As expected, the Wald method has the smallest confidence interval width of 4.46, which we believe is not an anomaly since this finding aligns with the simulation results.
 
  Finally, for all the 7 endpoints we have 36 wins, 16 losses, and 32 ties in the UDCA group with a p-value of 0.006 for our proposed test statistic and 0.003 for the Pocock et al. test statistic. The additional endpoints help break ties in large number of matched pairs, improve the statistical inference, and lead to tighter confidence intervals. Across the different methods, we also see similar trend in the confidence intervals for the net benefit and the win ratio as we found looking at only death and transplant. The confidence interval for the net benefit are fairly similar. For the win ratio, the Pocock et al. interval is the widest (width = 7.77), while the Wald method has the most narrow  (2.66). Both MOVER Wilson and MOVER AC methods as well as the Wald-LT have smaller confidence intervals (resp. 2.78, 2.81, and 2.80) compared to the interval of the Fieller's method  (3.32). 
 
Overall,  in the three examples, the confidence interval estimation methods considered for the net benefit provide similar results. For the win ratio, the results are similar  only  when the sample size is large or when the proportion of wins is adequately bounded away from 0. Otherwise, the Wald,  Fieller's and Pocock et al. method can yield unreliable results. Although the Wald-LT method often results in narrow confidence interval, the interval sensitive to the values of the proportions $p_w$ or $p_l$ The MOVER Wilson and AC as well as the Wald-LT provide more meaningful and realistic results, without any boundary anomaly or unnecessary wide confidence intervals, regardless of the sample size of the data we used or the proportion of wins we obtained.

\section{Conclusion}\label{sec:conclusion}
The use of large-sample confidence intervals for the difference and ratio of proportions comes with some caveats and conditions, which the current literature on the net benefit or the win ratio has not yet mentioned nor addressed. Fortuitously, in most of the examples and simulations considered so far, authors did not run into or were not aware of the deficiencies and inappropriate behaviors (e.g., non-conservatism and boundary violations) these large-sample confidence intervals have at or near the probability boundaries of 0 and 1 or when the sample size is small.\cite{agresti1998approximate,brown2001interval,  brown2005confidence}

  In this paper, we have made two major contributions in the statistical inference of the matched win ratio and the matched net benefit. First, we show that the test statistic  proposed by Pocock et al.\cite{pocock2011win} has serious limitations  to test for null hypothesis, especially under small sample size,  since it does not control for the type I error. We demonstrated that the related variance estimation for the matched win ratio proposed by Pocock et al.\cite{pocock2011win} is over-optimistic and thus may lead to unreliable results, especially when the total number of matched pairs is small. We thus proposed a different, yet simple test statistic instead and a formula for sample size calculation. The proposed test statistic outperforms the Pocock et al. test statistic: it adequately controls the type error and do not overestimate power. 

Second,  we should bear in mind that the confidence interval estimations for both the net benefit and the win ratio rely on the large-sample approximation. However, such large-sample methods to estimate the confidence interval for difference and ratio of proportions have been shown to behave badly when the proportions are near 0 or 1 or when the sample size is small. In this paper,   we have focused on  alternative methods to estimate the confidence intervals for the matched win ratio and the matched net benefit using the method of variance estimates recovery (MOVER). The proposed method  uses the idea that a confidence interval of a function of two proportions can be obtained from the individual confidence intervals of each proportion determined separately. 

The proposed MOVER confidence intervals thus incorporate  separate confidence intervals for individual proportions based on   Agresti-Coull (AC) and the Wilson single-proportion methods to construct confidence intervals the matched net benefit and matched win ratio. We then compared the proposed methods to the traditional large-sample Wald confidence estimation methods. 
For the matched net benefit, we compare the MOVER to the large-sample Wald confidence interval estimation method for the difference of proportions. For the win ratio, the proposed methods were compared to 4 other methods: the large-sample Wald method, the large-sample  Wald method based on a logarithm transformation, Pocock et al. method, and the Fieller's Theorem method.
 
  We have demonstrated through extensive simulation studies and data examples that  all the  methods perform reasonably well and produce fairly similar confidence intervals only when the sample size is large and the proportions of wins or losses are not extreme, i.e. near 0 or 1. However, the results  can be drastically different form each other when the sample size is small or when the proportions of wins or losses are extreme, with large-sample Wald methods and the Fieller's Theorem leading to aberrant and anomalous behaviors. 


Overall, our results demonstrate that the proposed MOVER methods outperform their competitors. They provide adequate coverage probability at the pre-specified nominal level, are slightly conservative (i.e., close or invariably at least equal to $1-\alpha$),\cite{newcombe1998two} have shorter confidence interval widths, and have good boundary-respecting properties, regardless of the sample size or the proportions of wins and losses. As such, they have a better coverage-width trade-off.
We, therefore,  recommend the use of the MOVER AC and Wilson methods to estimate the confidence intervals for  the matched net benefit and the matched win ratio.
	\section*{Acknowledgments}

We thank Dr. Sean O'Brien for carefully reading the manuscript and providing insightful and thoughtful suggestions and comments, which substantially improved the presentation of
the article.
		\appendix
\newcounter{Appendix}[section]
\numberwithin{equation}{section}
\renewcommand\theequation{\Alph{section}.\arabic{equation}}
\renewcommand\theequation{\Alph{section}.\arabic{table}}
\numberwithin{table}{section}
\numberwithin{figure}{section}
\section{Technical proofs}
\subsection{Test statistic}\label{appendix:test}
We show that the test statistic $Z_m$ is exactly equivalent to the large-sample test statistic $Z$ derived in the previous section by the equation \eqref{eq:test_binom}, using the normal approximation to the conditional binomial distribution \eqref{eq:binomial}.  
\begin{align*}
	Z_m&=\frac{p_w-p_l}{\sqrt{(p_w+p_l)/N}}=\displaystyle\frac{p_w-\frac{1}{2}(p_w+p_l)}{\frac{1}{2}\sqrt{(p_w+p_l)/N}}\\
	& =\displaystyle\frac{p_w/(p_w+p_l)-\frac{1}{2}}{\displaystyle\frac{\displaystyle\sqrt{(p_w+p_l)/N}}{2(p_w+p_l)}} 
	=\displaystyle\frac{Q_w-\frac{1}{2}}{\sqrt{\displaystyle\frac{1}{4N(p_w+p_l)}}}\\
	&  =\displaystyle\left(Q_w-\frac{1}{2}\right)\sqrt{4(N_w+N_l)}
\end{align*} 
which is equal to $Z$ as in the equation \eqref{eq:test_binom} of Section \ref{sec:pocock_short}.

Finally, if we replace $\displaystyle p_w$ and $\displaystyle p_l$ by   $\displaystyle p_w=\frac{N_w}{N}$  and $\displaystyle p_l=\frac{N_l}{N}$ in  equation $Z_m$, we have $Q_w=p_w/(p_w+p_l)=N_w/(N_w+N_l)$. We can write the test statistic $Z=Z_m$ as 
\begin{align*}
	Z&=\frac{N_w-N_l}{\sqrt{(N_w+N_l)}} 
\end{align*} 
The above equation indicates that the number of ties $N_t$ does not contribute directly to and has a negligible impact on the significance of the test statistic $Z$. Nevertheless, ties may have an impact on its effect size both in term of magnitude and precision. \cite{agresti2004effects} 

\subsection{Variance Estimation for the win ratio}\label{appendix:var_wr}
Consider $\displaystyle R_w=\frac{N_w}{N_l}=\frac{\ p_w}{ \ p_l}$, we can apply the multivariate delta method to the function $\displaystyle f(x,y,z)=\frac{x}{y}$ to calculate the variance of $  R_w.$\\
Since $\displaystyle \frac{\partial}{\partial x}f(x,y,z)=\frac{1}{y},$  $\displaystyle \frac{\partial}{\partial y}f(x,y,z)=\frac{-x}{y^2}$, and $\displaystyle \frac{\partial}{\partial z}f(x,y,z)=0,$ we can use  the delta method of function of random vector (see for instance, Agresti\cite{agresti2013categorical})
to calculate the variance of $R_w=f( p_w, p_w,  p_t).$
\begin{align*}
\begin{pmatrix} \displaystyle\frac{1}{\pi_l},&  -\displaystyle\frac{\pi_w}{\pi_l^2},& 0\end{pmatrix}
\begin{pmatrix}
\pi_w(1-\pi_w)/N & -\pi_w\pi_l/N & -\pi_w\pi_t/N \\\\
-\pi_w\pi_l/N & \pi_l(1-\pi_l)/N & -\pi_l\pi_t/N \\\\
-\pi_w\pi_t/N  & -\pi_l\pi_t/N & \pi_t(1-\pi_t)/N
\end{pmatrix}\begin{pmatrix}\displaystyle \frac{1}{\pi_l}\vspace{.2cm}\\  -\displaystyle\frac{\pi_w}{\pi_l^2}\vspace{.2cm}\\0\end{pmatrix}=\displaystyle \frac{1}{N}\begin{pmatrix}
\displaystyle \frac{\pi_w}{\pi_l} ,& -\displaystyle \frac{\pi_w}{\pi_l}  ,& 0 \\
\end{pmatrix}\begin{pmatrix}\displaystyle \frac{1}{\pi_l}\vspace{.2cm}\\  -\displaystyle\frac{\pi_w}{\pi_l^2}\vspace{.2cm}\\0\end{pmatrix}=\displaystyle\frac{\pi_w(\pi_w+\pi_l)}{N\pi_l^3}
\end{align*}
Therefore, $\displaystyle Var(R_w)=\frac{\pi_w(\pi_w+\pi_l)}{N\pi_l^3}.$
\subsection{Variance Estimation for the log($R_w$)}\label{appendix:var_logwr}
This time, we consider the function $\displaystyle g(x,y,z)=\log\left( \frac{x}{y}\right) =\log(x)-\log(y).$ The partial derivatives of $g$ are $\displaystyle \frac{\partial}{\partial x}g(x,y)=\frac{1}{x}$,    $\displaystyle \frac{\partial}{\partial y}g(x,y,z)=-\frac{1}{y}$, and $\displaystyle \frac{\partial}{\partial z}g(x,y,z)=0.$ 
Applying the  delta method of function of random vector to $log(R_w)=g(p_w, p_w, p_t)$, we have 
\begin{align*}
\begin{pmatrix} \displaystyle\frac{1}{\pi_w},&  -\displaystyle\frac{1}{\pi_l},& 0\end{pmatrix}
\begin{pmatrix}
\pi_w(1-\pi_w)/N & -\pi_w\pi_l/N & -\pi_w\pi_t/N \\\\
-\pi_w\pi_l/N & \pi_l(1-\pi_l)/N & -\pi_l\pi_t/N \\\\
-\pi_w\pi_t/N  & -\pi_l\pi_t/N & \pi_t(1-\pi_t)/N
\end{pmatrix}\begin{pmatrix}\displaystyle \frac{1}{\pi_w}\vspace{.2cm}\\  -\displaystyle\frac{1}{\pi_l }\vspace{.2cm}\\0\end{pmatrix}&=\displaystyle \frac{1}{N}\begin{pmatrix}
\displaystyle 1,& -1,& 0 
\end{pmatrix}\begin{pmatrix}\displaystyle \frac{1}{\pi_w}\vspace{.2cm}\\  -\displaystyle\frac{1}{\pi_l }\vspace{.2cm}\\0\end{pmatrix}=\displaystyle \frac{1}{N} \displaystyle\left( \frac{1}{\pi_w}+\frac{1}{ \pi_l}\right) 
\end{align*}
Therefore, $\displaystyle Var(\log\left( R_w)\right) =\displaystyle \frac{1}{N} \displaystyle\left( \frac{1}{\pi_w}+\frac{1}{ \pi_l}\right).$

\subsection{Variance  for $Y=p_w-Rp_l$}\label{appendix:var_diffwr}
Let $Y=\pi_w-Rp_l$ and consider the function $h$  defined by $\displaystyle h(x,y,z)=x-Ry,$ where $R=\displaystyle\frac{\pi_w}{\pi_l}$. 
The partial derivatives of $h$ are  $\displaystyle \frac{\partial}{\partial x}h(x,y)=1$,    $\displaystyle \frac{\partial}{\partial y}h(x,y,z)=-R$, and $\displaystyle \frac{\partial}{\partial z}h(x,y,z)=0.$ \\
The  delta method applied to $Y=h(p_w,  p_w, p_t)$ gives the variance $Var(Y)$ as
\begin{align*}
&\begin{pmatrix} 1,&  -R,& 0\end{pmatrix}
\begin{pmatrix}
\pi_w(1-\pi_w)/N & -\pi_w\pi_l/N & -\pi_w\pi_t/N \\\\
-\pi_w\pi_l/N & \pi_l(1-\pi_l)/N & -\pi_l\pi_t/N \\\\
-\pi_w\pi_t/N  & -\pi_l\pi_t/N & \pi_t(1-\pi_t)/N
\end{pmatrix}\begin{pmatrix}1\vspace{.2cm}\\  -R\vspace{.2cm}\\0\end{pmatrix}\\
&=\displaystyle \frac{1}{N}\begin{pmatrix}
\displaystyle \pi_w(1-\pi_w)+R\pi_w\pi_l,& -\pi_w\pi_l-R\pi_l(1-\pi_l),&  -\pi_w\pi_t-R\pi_l\pi_t 
\end{pmatrix}\begin{pmatrix}1\vspace{.2cm}\\  -R\vspace{.2cm}\\0\end{pmatrix}\\
&=\displaystyle \frac{1}{N} \displaystyle\left[ \pi_w(1-\pi_w)+2R\pi_w\pi_l+R^2\pi_l(1-\pi_l)\right]. 
\end{align*}
\subsection{Proof that  $B^2-AC\geq 0$}
\begin{align*}
B^2-AC&=p_w^2p_l^2\left( N+z_{\frac{\alpha}{2}}^2\right)^2\\
&-(Np_l^2-z_{\frac{\alpha}{2}}^2p_l(1-p_l))(Np_w^2-z_{\frac{\alpha}{2}}^2p_w(1-p_w))\\
&=p_w^2p_l^2(N^2+z_{\frac{\alpha}{2}}^4+2Nz_{\frac{\alpha}{2}}^2)-N^2p_l^2p_w^2+Np_l^2z_{\frac{\alpha}{2}}^2p_w(1-p_w)\\
&+Np_w^2z_{\frac{\alpha}{2}}^2p_l(1-p_l)+z_{\frac{\alpha}{2}}^4p_w(1-p_w)p_l(1-p_l)\\
&=Np_wp_lz_{\frac{\alpha}{2}}^2\left[  p_l(1-p_w)+p_w(1-p_l)+2p_wp_l\right]\\
& ~+ z_{\frac{\alpha}{2}}^4p_wp_l[p_wp_l+(1-p_w)(1-p_l)]\\
&=p_wp_lz_{\frac{\alpha}{2}}^2\left[ N(p_l+p_w+2p_wp_l)+ z_{\frac{\alpha}{2}}^2(2p_wp_l+p_t)\right] \geq 0
\end{align*}

\section{Simulation results}
Tables of the results for the different simulation studies conducted in the main paper are presented in this section.
\subsection{Type I error and power calculation}
In this section, we present the Tables for type I error (Table \ref{tab:typeIerror}) and power (Table \ref{tab:power}) of the tests statistic $Z_P,$ and $Z$ (given, respectively, by \eqref{eq:test_binomial} and \eqref{eq:test_binom}).
\begin{table}[ht!]
	\caption{Type I error estimation for the Pocock et al. and our proposed tests statistic}\label{tab:typeIerror}
	\centering
	\begin{threeparttable}
		\begin{tabular}{clccccc}
			\toprule
			& & \multicolumn{5}{c}{Sample size} \\\cmidrule(lr){3-7}
			$p_w $ & Test & 30 & 40 & 50 & 100 & 200\\
			\midrule
			\multirow{2}{*}{$0.1$}
			& $Z_P$ & 0.16 & 0.13 & 0.11 & 0.07 & 0.05 \\ 
			& $Z$  & 0.05 & 0.05 & 0.04 & 0.05 & 0.04\\ 
			\midrule
			\multirow{2}{*}{$0.2$}
			& $Z_P$  & 0.11 & 0.08 & 0.07 & 0.06 & 0.06 \\ 
			& $Z$    & 0.05 & 0.05 & 0.05 & 0.05 & 0.05 \\ 
			\midrule
			\multirow{2}{*}{$0.3$}
			& $Z_P$  & 0.07 & 0.06 & 0.06 & 0.06 & 0.05 \\
			& $Z$  & 0.05& 0.05 & 0.05 & 0.05 & 0.05 \\
			\midrule
			\multirow{2}{*}{$0.4$} 
			& $Z_P$ & 0.07 & 0.06 & 0.06 & 	0.05 & 	0.05\\
			& $Z$   & 0.07 & 0.05 & 0.05 & 	0.05 & 	0.05\\
			\midrule
			\multirow{2}{*}{$0.5$}
			& $Z_P$ & 0.05 & 0.09 & 0.07 & 0.06 & 0.06\\
			& $Z$   & 0.04 & 0.04 & 0.07 & 0.06 & 0.06\\	
			\bottomrule
			
		\end{tabular}
		\begin{tablenotes}\footnotesize
			\item Tests statistic: $Z_P$: Pocock et al.'s \eqref{eq:test_binomial};  $Z$: proposed \eqref{eq:test_binom}.
		\end{tablenotes}
	\end{threeparttable}
\end{table}

\begin{table}[]
	\caption{Estimated empirical power for the Pocock et al and the proposed tests statistic}\label{tab:power}
	\centering
	\begin{threeparttable}
		\begin{tabular}{clcccccccc}
			\toprule
			& & \multicolumn{8}{c}{Sample Size} \\
			\cmidrule(lr){3-10}
			$p_w$ & Test & 30 & 40 & 50 & 60 & 70 & 80 & 90 & 100\\
			\midrule
			\multirow{2}{*}{$0.4$}
			& $Z_P$& 0.12 & 0.13 & 0.15 & 0.18 & 0.20 & 0.21 & 0.22 & 0.23 \\
			& $Z$ & 0.11 & 0.11 & 0.14 & 0.15 & 0.17 & 0.18 &  0.20 & 0.22\\  
			\midrule
			\multirow{2}{*}{$0.5$}
			& $Z_P$& 0.27 & 0.32 & 0.39 & 0.44 & 0.49 & 0.54 & 0.57 & 0.62\\
			& $Z$& 0.24 & 0.30 & 0.35 & 0.41 & 0.45 & 0.52 & 0.56 & 0.61\\ 
			\midrule
			\multirow{2}{*}{$0.6$}
			& $Z_P$& 0.46 & 0.54 & 	0.66 & 0.72 & 0.78 & 0.83 & 0.87 & 0.91\\
			& $Z$ & 0.42 & 0.51 & 0.62 & 0.70 &	0.77 & 0.82	& 0.87 & 0.90\\
			\bottomrule	
		\end{tabular}
		\begin{tablenotes}\footnotesize
			\item Tests statistic: $Z_P$: Pocock et al.'s \eqref{eq:test_binomial};   $Z$: proposed \eqref{eq:test_binom}.
		\end{tablenotes}
	\end{threeparttable}
\end{table}
\newpage
\subsection{Estimated coverage probability and average width of the confidence interval}
\begin{table}[ht!]
	\caption{Estimated coverage probability of the 95\% confidence interval  of the net benefit.}\label{tab:coverage_nb}
	\centering
	\begin{threeparttable}
		\begin{tabular}{crcccccccccccccccccccccccccccccccccccccccccccccccccccccc}
			\toprule
			&  & \multicolumn{3}{c}{$N=30$} &  \multicolumn{3}{c}{$N=50$}\\\cmidrule(lr){3-5}\cmidrule(lr){7-9}
			
			NB\multirow{9}{*}{0.25}&$P_t$& Wald & AC & Wilson  &&  Wald & AC & Wilson\\ \midrule
			&   0.1  & 0.94 & 0.96 & 0.96   && 0.95 & 0.96 & 0.96 \\
			&   0.15 & 0.94 & 0.96 & 0.96   && 0.94 & 0.95 & 0.95 \\ 
			&   0.2  & 0.94 & 0.96 & 0.96   && 0.94 & 0.95 & 0.95 \\ 
			&   0.25 & 0.93 & 0.95 & 0.95   && 0.94 & 0.95 & 0.95 \\
			&   0.3  & 0.93 & 0.95 & 0.95   && 0.94 & 0.95 & 0.94 \\
			&   0.35 & 0.93 & 0.95 & 0.95   && 0.93 & 0.95 & 0.94 \\
			&   0.4  & 0.93 & 0.95 & 0.95   && 0.94 & 0.94 & 0.95 \\
			&   0.45 & 0.93 & 0.96 & 0.96   && 0.94 & 0.95 & 0.95 \\
			&   0.5  & 0.94 & 0.96 & 0.96   && 0.94 & 0.95 & 0.95 \\
			\addlinespace
			\multirow{9}{*}{0.375}
			& 0.1  & 0.93 & 0.95 &  0.95   && 0.94 & 0.96 & 0.96  \\
			& 0.15 & 0.93 & 0.95 &  0.95   && 0.94 & 0.95 & 0.95  \\
			& 0.2  & 0.94 & 0.96 &  0.95   && 0.94 & 0.95 & 0.95  \\
			& 0.25 & 0.93 & 0.95 &  0.95   && 0.94 & 0.95 & 0.95  \\
			& 0.3  & 0.93 & 0.95 &  0.95   && 0.94 & 0.95 & 0.95  \\
			& 0.35& 0.93  & 0.96 &  0.95   && 0.94 & 0.96 & 0.96  \\
			& 0.4  & 0.92 & 0.96 &  0.95   && 0.94 & 0.95 & 0.95  \\
			& 0.45 & 0.93 & 0.96 &  0.96   && 0.94 & 0.95 & 0.95  \\
			& 0.5  & 0.93 & 0.96 &  0.96   && 0.94 & 0.96 & 0.96  \\
			\addlinespace\multirow{9}{*}{0.50}
			&  0.1	&  0.92 & 0.95 & 0.95 && 0.93 &   0.95 &   0.95    \\
			&  0.15 &  0.93 & 0.95 & 0.94 && 0.94 &   0.95 &   0.95    \\
			&  0.2	&  0.93 & 0.95 & 0.95 && 0.93 &   0.95 &   0.95    \\
			&  0.25 &  0.92 & 0.94 & 0.94 && 0.94 &   0.96 &   0.95    \\
			&  0.3	&  0.93 & 0.95 & 0.95 && 0.94 &   0.95 &   0.95    \\
			&  0.35 &  0.94 & 0.95 & 0.95 && 0.93 &   0.96 &   0.95    \\
			&  0.4	&  0.94 & 0.95 & 0.95 && 0.94 &   0.96 &   0.95    \\
			&  0.45 &  0.95 & 0.96 & 0.96 && 0.94 &   0.96 &   0.95    \\
			&  0.5	&  0.95 & 0.97 & 0.97 && 0.94 &   0.96 &   0.95    \\
			\bottomrule
		\end{tabular}
		\begin{tablenotes}\footnotesize
			\item $P_t$: probability of ties. 
		\end{tablenotes}
	\end{threeparttable}
\end{table}

\begin{table}[h!]
	\caption{ Average width of the 95\% confidence interval  of the net benefit}\label{tab:width_nb}
	\centering
	\begin{threeparttable}
		\begin{tabular}{crcccccccccccccccccccccccccccccccccccccccccccccccccccccc}
			\toprule
			&  & \multicolumn{3}{c}{$N=30$} &  \multicolumn{3}{c}{$N=50$}\\\cmidrule(lr){3-5}\cmidrule(lr){7-9}
			
			NB\multirow{9}{*}{0.25}&$P_t$& Wald & AC & Wilson  &&  Wald & AC & Wilson\\ \midrule
			&   0.1  & 0.64 & 0.61 & 0.61   && 0.50 & 0.49 & 0.49 \\
			&   0.15 & 0.62 & 0.60 & 0.59   && 0.49 & 0.47 & 0.47 \\ 
			&   0.2  & 0.60 & 0.58 & 0.57   && 0.47 & 0.46 & 0.46 \\ 
			&   0.25 & 0.58 & 0.56 & 0.55   && 0.45 & 0.44 & 0.44 \\
			&   0.3  & 0.56 & 0.54 & 0.53   && 0.43 & 0.43 & 0.42 \\
			&   0.35 & 0.54 & 0.52 & 0.51   && 0.42 & 0.41 & 0.41 \\
			&   0.4  & 0.51 & 0.50 & 0.50   && 0.40 & 0.40 & 0.39 \\
			&   0.45 & 0.50 & 0.48 & 0.48   && 0.38 & 0.38 & 0.38 \\
			&   0.5  & 0.46 & 0.47 & 0.46   && 0.36 & 0.36 & 0.36 \\
			\addlinespace
			\multirow{9}{*}{0.375}
			& 0.1  & 0.61 & 0.59 &  0.58   && 0.48 & 0.47 & 0.46  \\
			& 0.15 & 0.59 & 0.57 &  0.56   && 0.46 & 0.45 & 0.44  \\
			& 0.2  & 0.57 & 0.55 &  0.55   && 0.44 & 0.44 & 0.43  \\
			& 0.25 & 0.55 & 0.53 &  0.53   && 0.43 & 0.42 & 0.42  \\
			& 0.3  & 0.52 & 0.51 &  0.50   && 0.41 & 0.41 & 0.40  \\
			& 0.35 & 0.50 & 0.50 &  0.49   && 0.39 & 0.39 & 0.38  \\
			& 0.4  & 0.47 & 0.48 &  0.47   && 0.37 & 0.37 & 0.37  \\
			& 0.45 & 0.45 & 0.46 &  0.45   && 0.35 & 0.36 & 0.35  \\
			& 0.5  & 0.42 & 0.45 &  0.43   && 0.33 & 0.34 & 0.33  \\
			\addlinespace\multirow{9}{*}{0.50}
			&  0.1	&  0.56 & 0.55 & 0.54 && 0.44 &   0.44 &   0.43    \\
			&  0.15 &  0.54 & 0.53 & 0.52 && 0.42 &   0.42 &   0.41    \\
			&  0.2	&  0.52 & 0.51 & 0.50 && 0.40 &   0.40 &   0.40    \\
			&  0.25 &  0.49 & 0.50 & 0.48 && 0.39 &   0.39 &   0.38    \\
			&  0.3	&  0.47 & 0.47 & 0.46 && 0.37 &   0.37 &   0.36    \\
			&  0.35 &  0.44 & 0.45 & 0.44 && 0.34 &   0.35 &   0.34    \\
			&  0.4	&  0.41 & 0.43 & 0.42 && 0.32 &   0.33 &   0.32    \\
			&  0.45 &  0.38 & 0.41 & 0.39 && 0.30 &   0.31 &   0.30    \\
			&  0.5	&  0.35 & 0.38 & 0.37 && 0.27 &   0.29 &   0.28    \\
			\bottomrule
		\end{tabular}
		\begin{tablenotes}\footnotesize
			\item $P_t$: probability of ties. 
		\end{tablenotes}
	\end{threeparttable}
\end{table}

\begin{table}[h!]
	\caption{ Estimated coverge probability of the 95\% confidence interval  of the win ratio}\label{tab:coverage_wr}
	\centering
	\begin{threeparttable}
		\begin{tabular}{crcccccccccccccccccccccccccccccccccccccccccccccccccccccc}
			\toprule
			&  & \multicolumn{6}{c}{$N=30$} &  \multicolumn{6}{c}{$N=50$}\\\cmidrule(lr){3-8}\cmidrule(lr){10-15}
			
			WR\multirow{9}{*}{1}&$P_t$& Pocock & Wald & Wald-LT & Fieller & AC & Wilson  && Pocock & Wald & Wald-LT & Fieller & AC & Wilson\\ \midrule
			&   0.1  & 0.94 & 0.93 & 0.95 & 0.93 & 0.94 & 0.94   && 0.94 & 0.96 & 0.95  & 0.94 & 0.95 & 0.95 \\
			&   0.15 & 0.94 & 0.93 & 0.96 & 0.93 & 0.95 & 0.95   && 0.94 & 0.93 & 0.96  & 0.94 & 0.95 & 0.95 \\ 
			&   0.2  & 0.94 & 0.93 & 0.96 & 0.94 & 0.95 & 0.95   && 0.93 & 0.93 & 0.95  & 0.93 & 0.95 & 0.95 \\ 
			&   0.25 & 0.93 & 0.92 & 0.96 & 0.93 & 0.95 &	0.95  && 0.94 & 0.94 & 0.95  & 0.94 & 0.95 & 0.95 \\
			&   0.3  & 0.93 & 0.92 & 0.96 & 0.93 & 0.95 & 0.95   && 0.94 & 0.93 & 0.95  & 0.94 & 0.95 & 0.95 \\
			&   0.35 & 0.93 & 0.92 & 0.96 & 0.93 & 0.94 & 0.94   && 0.94 & 0.93 & 0.95  & 0.94 & 0.95 & 0.95 \\
			&   0.4  & 0.92 & 0.91 & 0.96 & 0.93 & 0.95 & 0.95   && 0.93 & 0.94 & 0.95  & 0.93 & 0.95 & 0.94 \\
			&   0.45 & 0.92 & 0.91 & 0.96 & 0.93 & 0.95 & 0.95   && 0.94 & 0.09 & 0.96  & 0.94 & 0.95 & 0.95 \\
			&   0.5  & 0.92 & 0.91 & 0.96 & 0.92 & 0.96 & 0.96   && 0.93 & 0.10 & 0.95  & 0.94 & 0.95 & 0.95 \\
			\addlinespace
			\multirow{9}{*}{1.5}
			& 0.1  & 0.93 & 0.94 & 0.95 &  0.93 & 0.95 & 0.95   && 0.94 & 0.94 & 0.95  & 0.94 & 0.95  &  0.95  \\
			& 0.15 & 0.93 & 0.94 & 0.95 &  0.93 & 0.95 & 0.95   && 0.94 & 0.94 & 0.95  & 0.94 & 0.95  &  0.95  \\
			& 0.2  & 0.93 & 0.94 & 0.95 &  0.93 & 0.95 & 0.95   && 0.94 & 0.93 & 0.96  & 0.94 & 0.95  &  0.95  \\
			& 0.25 & 0.93 & 0.93 & 0.96 &  0.93 & 0.96 & 0.95   && 0.94 & 0.93 & 0.95  & 0.94 & 0.95  &  0.95  \\
			& 0.3  & 0.93 & 0.93 & 0.95 &  0.93 & 0.95 & 0.95   && 0.94 & 0.93 & 0.95  & 0.94 & 0.95  &  0.95  \\
			& 0.35& 0.92 & 0.92 & 0.96 &  0.93 & 0.95 & 0.95   && 0.94 & 0.93 & 0.96  & 0.94 & 0.95  &  0.95  \\
			& 0.4  & 0.92 & 0.92 & 0.96  & 0.91 & 0.95 & 0.95   && 0.93 & 0.93 & 0.95  & 0.93 & 0.94  &  0.94  \\
			& 0.45 & 0.91 & 0.92 & 0.96 &  0.89 & 0.95 & 0.95   && 0.93 & 0.94 & 0.96  & 0.94 & 0.95  &  0.95  \\
			& 0.5  & 0.89 & 0.91 &  0.97 & 0.85 & 0.96 & 0.95   && 0.93 & 0.94 & 0.95  & 0.94 & 0.95  &  0.95  \\
			\addlinespace\multirow{9}{*}{2}
			&  0.1	 &  0.92 &  0.93 & 0.96 & 0.92 &  0.95 &  0.94  && 0.94 & 0.93 & 0.95  & 0.94 &  0.95  &  0.95    \\
			&  0.15 &  0.93 &  0.93 & 0.97 & 0.93 &  0.95 &  0.95  && 0.93 & 0.93 & 0.95  & 0.93 &  0.95  &  0.95    \\
			&  0.2	 &  0.92 &  0.93 & 0.96 & 0.93 &  0.96 &  0.95  && 0.93 & 0.93 & 0.95  & 0.93 &  0.95  &  0.94    \\
			&  0.25 &  0.92 &  0.92 & 0.96 & 0.92 &  0.96 &  0.95  && 0.93 & 0.93 & 0.95  & 0.94 &  0.95  &  0.95    \\
			&  0.3	 &  0.91 &  0.91 & 0.96 & 0.92 &  0.96 &  0.95  && 0.93 & 0.93 & 0.96  & 0.94 &  0.95  &  0.95      \\
			&  0.35 &  0.89 & 	0.92 & 0.97	& 0.89 &  0.95 &  0.95  && 0.93 & 0.93 & 0.95  & 0.94 &  0.95  &  0.95     \\
			&  0.4	 &  0.86 &  0.92 & 0.96 & 0.85 &  0.95 &  0.95  && 0.93 & 0.93 & 0.96  & 0.94 &  0.95  &  0.95     \\
			&  0.45 &  0.83 & 	0.91 & 0.97 & 0.81 &  0.95 &  0.95  && 0.93 & 0.93 & 0.96  & 0.94 &  0.95  &  0.95      \\
			&  0.5	 &  0.81 & 	0.91 & 0.97 & 0.75 &  0.94 &  0.96  && 0.92 & 0.92 & 0.96  & 0.93 &  0.96  &  0.94    	 \\
			\bottomrule
		\end{tabular}
		\begin{tablenotes}\footnotesize
			\item $P_t$: probability of ties; Wald-LT: Wald confidence interval based on a logarithm transformation. 
		\end{tablenotes}
	\end{threeparttable}
\end{table}

\begin{table}[h!]
	\caption{ Average width of the 95\% confidence interval of the win ratio.} \label{tab:width_wr}
	\centering
	\begin{threeparttable}
		\begin{tabular}{crcccccccccccccccccccccccccccccccccccccccccccccccccccccc}
			\toprule
			&  & \multicolumn{6}{c}{$N=30$} &  \multicolumn{6}{c}{$N=50$}\\\cmidrule(lr){3-8}\cmidrule(lr){10-15}
			
			WR\multirow{9}{*}{1}&$P_t$& Pocock & Wald & Wald-LT & Fieller & AC & Wilson  && Pocock & Wald & Wald-LT & Fieller & AC & Wilson\\ \midrule
			&   0.1  & 2.09 & 1.68 & 1.85 & 2.12 & 1.81 & 1.80   && 1.39 & 1.25 & 1.32 & 1.39 & 1.30 & 1.30 \\
			&   0.15 & 2.21 & 1.74 & 1.93 & 2.26 & 1.88 & 1.87   && 1.45 & 1.29 & 1.37 & 1.46 & 1.35 & 1.35 \\ 
			&   0.2  & 2.43 & 1.83 & 2.05 & 2.53 & 2.01 & 1.99   && 1.51 & 1.33 & 1.42 & 1.53 & 1.40 & 1.40 \\ 
			&   0.25 & 2.57 & 1.88 & 2.12 & 2.71 & 2.09 & 2.05   && 1.58 & 1.38 & 1.48 & 1.61 & 1.46 & 1.45 \\
			&   0.3  & 2.84 & 1.99 & 2.27 & 3.10 & 2.26 & 2.19   && 1.68 & 1.45 & 1.56 & 1.73 & 1.54 & 1.53 \\
			&   0.35 & 3.31 & 2.14 & 2.48 & 3.56 & 2.47 & 2.38   && 1.77 & 1.50 & 1.63 & 1.83 & 1.61 & 1.59 \\
			&   0.4  & 4.19 & 2.26 & 2.68 & 3.91 & 2.69 & 2.55   && 1.94 & 1.59 & 1.74 & 2.02 & 1.72 & 1.70 \\
			&   0.45 & 5.95 & 2.40 & 2.89 & 4.40 & 2.96 & 2.74   && 2.09 & 1.68 & 1.85 & 2.21 & 1.84 & 1.81 \\
			&   0.5  & 7.39 & 2.60 & 3.23 & 4.92 & 3.33 & 3.02   && 2.36 & 1.80 & 2.01 & 2.60 & 2.01 & 1.95 \\
			\addlinespace
			\multirow{9}{*}{1.5}
			& 0.1  & 4.02 & 2.66 & 2.97 &  4.12 & 2.94 & 2.88   && 2.31 & 1.95 & 2.07 & 2.33 & 2.05 & 2.04  \\
			& 0.15 & 4.45 & 2.80 & 3.15 &  4.64 & 3.14 & 3.05   && 2.45 & 2.03 & 2.17 & 2.47 & 2.15 & 2.13  \\
			& 0.2  & 4.76 & 2.98 & 3.41 &  5.06 & 3.39 & 3.29   && 2.60 & 2.06 & 2.25 & 2.65 & 2.23 & 2.21  \\
			& 0.25 & 5.15 & 3.12 & 3.62 &  3.12 & 3.64 & 3.48   && 2.78 & 2.18 & 2.36 & 2.85 & 2.35 & 2.31  \\
			& 0.3  & 5.72 & 3.23 & 3.78 &  6.17 & 3.89 & 3.62   && 2.99 & 2.29 & 2.49 & 3.10 & 2.49 & 2.44  \\
			& 0.35 & 6.09 & 3.54 & 4.26 &  3.54 & 4.26 & 4.04   && 3.16 & 2.36 & 2.58 & 3.30 & 2.58 & 2.53  \\
			& 0.4  & 8.32 & 3.86 & 4.77 &  7.40 & 4.92 & 4.48   && 3.67 & 2.54 & 2.81 & 4.01 & 2.85 & 2.74  \\
			& 0.45 &11.85 & 4.01 & 5.04 &  7.92 & 5.35 & 4.70   && 4.06 & 2.69 & 3.00 & 4.56 & 3.05 & 2.92  \\
			& 0.5  &15.66 & 4.31 & 5.58 &  8.13 & 6.15 & 5.15   && 4.79 & 2.96 & 3.40 & 5.74 & 3.47 & 3.28  \\
			\addlinespace\multirow{9}{*}{2}
			&  0.1	&  7.17 &  3.96 & 4.53 & 7.42 &  4.74 &  4.36  && 3.67 & 2.77 & 2.97 & 3.70 & 2.97 & 2.92 \\
			&  0.15 &  7.57 &  4.12 & 4.75 & 8.01 &  4.93 &  4.57  && 3.84 & 2.88 & 3.10 & 3.90 & 3.10 & 3.05 \\
			&  0.2	&  8.20 &  4.38 & 5.15 & 8.95 &  5.24 &  4.91  && 4.13 & 3.00 & 3.25 & 4.28 & 3.27 & 3.19 \\
			&  0.25 &  8.56 &  4.71 & 5.67 & 9.67 &  5.89 &  5.37  && 4.54 & 3.14 & 3.43 & 4.72 & 3.49 & 3.36 \\
			&  0.3	&  8.78 &  4.86 & 5.95 &10.02 &  6.22 &  5.61  && 4.99 & 3.27 & 3.60 & 5.30 & 3.67 & 3.52 \\
			&  0.35 &  9.64 &  5.27 & 6.59 &10.92 &  7.05 &  6.16  && 5.49 & 3.45 & 3.85 & 5.98 & 3.94 & 3.74 \\
			&  0.4	& 11.96 &  5.77 & 7.49 &10.87 &  7.77 &  6.90  && 6.30 & 3.70 & 4.17 & 7.16 & 4.45 & 4.05 \\
			&  0.45 & 16.61 &  6.05 & 8.01 &10.41 &  8.65 &  7.32  && 6.89 & 3.94 & 4.49 & 8.16 & 4.94 & 4.35 \\
			&  0.5	& 24.03 &  6.52 & 8.96 & 9.84 &  9.26 &  8.06  && 7.56 & 4.36 & 5.14 & 9.63 & 5.89 & 4.91 \\
			\bottomrule
		\end{tabular}
		\begin{tablenotes}\footnotesize
			\item $P_t$: probability of ties; Wald-LT: Wald confidence interval based on a logarithm transformation. 
		\end{tablenotes}
	\end{threeparttable}
\end{table}

\newpage

\bibliography{matched_WR_arxiv}

\end{document}